\documentclass[10pt,aps,prx,showpacs,superscriptaddress,reprint,citeautoscript]{revtex4-1}

\usepackage{amssymb,amsmath,amsfonts,bm} 
\usepackage{graphicx}				\usepackage{color} 
\usepackage{dcolumn}   
\usepackage{bm}        
\usepackage{url}

\usepackage{hyperref}
\usepackage{subfigure}

\def\eeq{\relax}
\def\beq#1#2\eeq{\begin{equation}\label{#1}#2\end{equation}}
\def\bal#1#2\eal{\begin{align}\label{#1}#2\end{align}}
\def\bse#1#2\ese{\begin{subequations}\label{#1}#2\end{subequations}}
\def\ba{\begin{aligned}}   \def\ea{\end{aligned}}

\def\XXint#1#2#3{{\setbox0=\hbox{$#1{#2#3}{\int}$}
\vcenter{\hbox{$#2#3$}}\kern-.5\wd0}}

\newcommand{\ii}{\ensuremath{\mathrm{i}}}

\DeclareMathOperator{\real}{Re}
\DeclareMathOperator{\imag}{Im}

\def\dd{\operatorname{d}} 

\begin{document}

\title{Non-symmetric flexural wave scattering and one-way extreme absorption
} 
\author{Andrew N. Norris}
\affiliation{Mechanical and Aerospace Engineering, Rutgers University, Piscataway, NJ 08854-8058 (USA)}
\author{Pawel Packo}
\affiliation{Department of Robotics and Mechatronics, AGH - University of Science and Technology,
Al. A. Mickiewicza 30, 30-059 Krakow, Poland}
\date{\today}

 \begin{abstract}
 
The possibility of  asymmetric absorption and reflection for flexural waves is demonstrated though analytical and numerical examples.  We focus on the 1D case of flexural motion of a beam  and consider combinations of point scatterers which together provide  asymmetric scattering.  The scatterers are attached damped oscillators characterized by effective impedances, analogous to effective configurations in 1D acoustic waveguides.  By selecting the impedances of a pair of closely spaced scatterers we show that it is possible to obtain almost total absorption for incidence on one side, with almost total reflection if incident from the other side.  The one-way absorption is illustrated through numerous examples of impedance pairs that satisfy the necessary conditions for zero reflectivity for incidence from one direction.   Examples of almost total and zero reflection for different incidences are examined in detail, showing the distinct wave dynamics of flexural waves as compared with acoustics.  

 \end{abstract}

\maketitle

\section{Introduction}\label{sec1}

{An isolated lumped element in an acoustic waveguides produces symmetric reflection for sound incident from either side. This is true for standard sub-wavelength scatterers such as a  side-branch Helmholtz resonator or a membrane stretched across the width of the waveguide.  However, by combining elements, e.g. a Helmholtz resonator (HR) and a  membrane in series, one can achieve asymmetric reflection depending on the direction of incidence. 
 Such a  combination of two or more point scatterers in a subwavelength configuration can be viewed as a new type of lumped element, called a Willis element 
\cite{Muhlestein2017a}.  Unlike the classical point scatterers the Willis element couples monopole  and dipole radiation which in turn leads to asymmetry in the scattering, while it can still be viewed as a sub-wavelength point scatterer.    The effective Willis parameters can be deduced from the scattering matrix elements \cite{Su2018a}.  
 An interesting  case of asymmetric reflection is unidirectional zero reflection \cite{Merkel2018} in which the reflection is zero for incidence from one direction but non-zero from the other. The extreme limit of this phenomenon is {\it one way total absorption} where unidirectional zero reflection is accompanied by  zero transmission.  The transmission must therefore be zero for incidence from both directions, as required by acoustical reciprocity. 
However, while total asymmetric acoustic  absorption implies zero symmetric transmission, the  reflection coefficients can differ as much as zero and unity in magnitude.  To the authors' knowledge, this extreme limit of 
one way total absorption has not yet been demonstrated for the simplest setup: 1D waveguides. }


%
%
%

The purpose of this paper is to show that   one way total absorption can be obtained for flexural waves.   Our analytical and numerical model is a 1D system  of flexural wave motion in a beam, for which,  by analogy with  the lumped elements in an acoustic waveguide, we consider closely spaced  translational point impedances.  These may be modeled as attached single degree of freedom damped oscillators which apply an effective point force to  the beam at the attachment point.  We do not consider rotational impedance elements which can  apply a moment \cite{Mead82}.  Through proper choice of the complex impedances, we demonstrate that two attached damped oscillators display the same quantitative wave effects as acoustic one way total absorption.  Specifically, reflection is zero for incidence from one side, while the reflection coefficient can be large, approaching unity in some cases, for waves incident from the opposite direction. 

The present problem is related to but fundamentally differs from the control of flexural waves in a beam using a passive  {\it tuned    vibration absorber} (TVA) \cite{Brennan1997,Brennan1999,El-Khatib2005}. 
A TVA,  modeled as a point translational impedance, can be used  to minimize transmission or to reduce the vibration at a specific frequency 
for a source that is either in the farfield  \cite{Brennan1999} or the near-field \cite{El-Khatib2005}.  The   term {\it vibration neutralizer} \cite{Brennan1997} rather than vibration  absorber is sometimes used  to signify that the purpose of the point attachment  is   to control vibration at a  particular frequency.   

Unlike a single TVA which  necessarily has symmetric scattering properties for incidence  from the left or the right, we consider two nearby impedances with the objective of maximizing the scattering asymmetry to obtain flexural wave one-way total absorption.  The design objective is quite distinct from that of the TVA in that we wish to make one reflection zero and other as close to unity as possible. 
Here we are only concerned with passive wave control.  We note that the reflection/transmission from two identical impedances was considered by \cite{Yang2016} 
where the effect of the spacing between the oscillators was found to be significant. 
However, the symmetric configuration gives the same reflection independent of the direction of incidence.  

 The outline of the paper is as follows. In Section \ref{sec2}  the governing equations are introduced and the solution is derived for scattering from two point impedances.    Necessary and sufficient conditions for one of the reflection coefficients to vanish are derived in Section \ref{sec3}.  It is shown that asymmetric reflection requires at least one of the oscillators must  be damped.  In
Section \ref{sec4} we show through numerous examples that one-way flexural reflection can be achieved from a wide variety of impedance pairs.     For instance, one may be purely real (undamped) and the other  imaginary (a pure damper), or both may be damped.  We find, surprisingly, that it is possible to achieve almost perfect one-way reflection (zero one way, unity the other).  This effect is explored in detail using asymptotic analysis in Section \ref{sec5}.  {Finally, it is shown in Section \ref{sec6} that 
almost perfect one way reflection is achievable with a unique pair of impedances in the frequency range $ka \approx (\pi, 2 \pi)$.}  The main results are summarized and conclusions are presented in Section \ref{sec7}. 


\section{Scattering by a cluster of point attachments}\label{sec2}

\subsection{General solution} 
The beam has bending stiffness $D$ $(=EI)$ and density $\rho '$ per unit length. 
Time harmonic motion $e^{-\ii \omega t}$ is assumed, so that the flexural wavenumber $k$ is defined by $k^4 = \omega^2 \rho '/D$. 
We assume there are $N$ point scatterers located at 
$ x_\alpha  $ 
with impedances $\mu_\alpha$, $\alpha = 1, 2 , \ldots, N$.  The total displacement $w$  satisfies 
\beq{-214}
D\big( \frac{\dd^4 w(x) }{\dd x^4}- k^4 w  \big) = \sum_{\alpha =1}^N 
\mu_\alpha w(x_\alpha) 
\delta(x-x_\alpha)  .
\eeq
The attachment impedance $\mu$ is modeled as single degree of freedom with  mass $M$, spring stiffness  $\kappa$  and damping coefficient $\nu$, {and defined as $\mu = f / w$, where $f$ denotes the driving force.  The impedance as used here is analogous to that for acoustics (pressure/velocity) although the two are not dimensionally equivalent.} Two possible configurations are 
\beq{-1}
\mu  = 
\begin{cases} 
\big( \frac 1{M\omega^2} - \frac 1{\kappa - \ii \omega \nu} \big)^{-1},
 & (a) ,
\\
M\omega^2 - \kappa + \ii \omega \nu , & (b)  .
\end{cases}
\eeq
In case (a) the mass is attached to the plate by a spring and damper in parallel \cite{Torrent2013}, also called a vibration neutralizer \cite{Brennan1999}.  
Model (b) assumes the mass is rigidly attached to the plate, and both are attached to a rigid foundation by the spring and damper in parallel \cite{Evans2007}.  An important limit is a  beam pinned at $x_\alpha$, $w(x_\alpha)=0$, which corresponds to $\mu \to \infty$. The (a) and (b) oscillators could also be attached in parallel, e.g.\ on either side of the beam, to give $\mu = \mu_a +\mu_b$.  The main point is that there is a wide range of achievable passive $\mu$ $(\imag \mu \ge 0)$. We take advantage of this adaptivity by exploring  the space of possible impedances in this paper. 

The solution 
is given by the incident  wave  $w_\text{in}(x)$ plus the displacement scattered
by all the particles, 
\beq{534}
w(x) = w_\text{in}(x) + \sum_{\beta = 1}^N G(x-x_\beta)  \mu_\beta w(x_\beta) 
\eeq
where the Green's function satisfies 
\beq{0333}
D\big( \frac{\dd^4 G(x) }{\dd x^4}- k^4 G(x) \big) = \delta (x) .
\eeq

We choose to normalize parameters so that the impedances and the Green's function are  nondimensional.  Define
\bse{5=5}
\bal{5=55}
m_\alpha &= \frac{ 2Dk^3}{\mu_\alpha}, 
\\
g(x) &= 2Dk^3 G(x) = \frac {\ii}{2} \big( e^{\ii k |x|} + \ii e^{-k|x|} \big), 
\eal
\ese
then \eqref{534} becomes
\beq{5}
w(x) = w_\text{in}(x) + \sum_{\beta = 1}^N g(x-x_\beta)  m_\beta^{-1} w(x_\beta) . 
\eeq
Setting $x=x_\alpha$ in \eqref{5} gives a linear system of $N$ equations which may be solved to give 
\beq{5-0}
w(x) = w_\text{in}(x) + \sum_{\alpha, \beta = 1}^N  g(x-x_\alpha) 
M_{\alpha\beta}^{-1} \, w_\text{in}(x_\beta)
\eeq
where $M_{\alpha\beta}^{-1}$ are the elements of  the inverse of the $N \times N$ matrix with elements 
\beq{5-1}
M_{\alpha\beta} = m_{\alpha} \delta_{\alpha\beta} -  g(x_\alpha -x_\beta)  .
\eeq

\subsection{Reflection, transmission and absorption coefficients}
We consider incidence from the left and right, $w_\text{in}^+$ and  $w_\text{in}^-$, 
respectively, 
\beq{033}
w_\text{in}^{\pm}(x) = e^{\pm \ii k x} . 
\eeq
The reflection coefficients $R_+$, $R_-$ and the single transmission coefficient $T$ are defined by 
\bse{44=}
\bal{5=32}
w(x) &= \begin{cases}
w_\text{in}^+ +R_+ e^{- \ii k x} , & x\to -\infty, 
\\
w_\text{in}^- +R_- e^{ \ii k x} , & x\to \infty  , 
\end{cases}
\\
w(x) &= T e^{\pm \ii k x} , \ \  x\to \pm \infty \ \text{for }\ w_\text{in}^{\pm}.
\eal
\ese
These follow from \eqref{5-0} as 
\bse{66-}
\bal{666}
R_{\pm} &= \frac {\ii}{2} \sum_{\alpha, \beta = 1}^N   
M_{\alpha\beta}^{-1}\,  e^{\pm \ii k (x_\alpha +x_\beta)} ,
\\
T&= 1+ \frac {\ii}{2} \sum_{\alpha, \beta = 1}^N   
M_{\alpha\beta}^{-1}\, e^{\ii k (x_\alpha -x_\beta)} .
\eal
\ese

{To quantify absorption, we define absorption coefficients for right and left incidence $\alpha^+$ and $\alpha^-$, respectively, as}
{
\beq{5-123a}
\alpha^\pm = 1 - |T|^2 - |R_\pm|^2.
\eeq
}

\subsection{Example: One and two scatterers}
It is useful to recall some of the features of a single translational impedance before we consider two attachments. 

\subsubsection{One scatterer}
For a single scatterer at $x=0$ we have $R_\pm = R = \frac{\ii}2 (m-g(0))^{-1}$ and $T = 1+R$. 
{A desired value of $T$ (or $R=T-1$) is obtained if 
\beq{566}
m = -\frac 12 + \frac{\ii T}{2(T-1)}
\ \ \Rightarrow \ \ w(0) = -\ii +(1+\ii )T . 
\eeq
Thus, for instance,  $w(0)=0$ for $m=0$ (infinite impedance $\mu$).  Even though the beam is pinned the rotation at $x=0$ is not constrained, and hence half of the incident energy is transmitted and half is reflected:
$|R|=|T|=1/\sqrt{2}$.   Zero transmission ($T=0$) is obtained  if $m=-\frac 12$, which is interpreted in terms of model (a) of Eq.\ \eqref{-1} by \cite{Brennan1999,El-Khatib2005}: the unique frequency at which $T$ vanishes is given by \cite[Eq.\ (7)]{Brennan1999} or \cite[Eq.\ (22)]{El-Khatib2005}. 
 In general, the range of possible values of $T$ for the single scatterer is restricted only by the requirement that the attachment is passive, i.e.\ $\imag m \le 0$ $\Leftrightarrow$
$\real T \ge |T|^2 \le 1$. }

{The full transmission or - equivalently - the zero reflection case, namely $|T| \rightarrow 1 \Leftrightarrow |R| \rightarrow 0$, from \eqref{566} corresponds to $m \rightarrow \infty \Leftrightarrow \mu \rightarrow 0$, so no scatterer. Also $w (x = 0) = e^{\ii k x} \vline_{x = 0} = 1$ as given by \eqref{566} with $T = 1$.}

\subsubsection{Two scatterers}

\begin{figure}[h!]
	\includegraphics[width=0.78\linewidth]{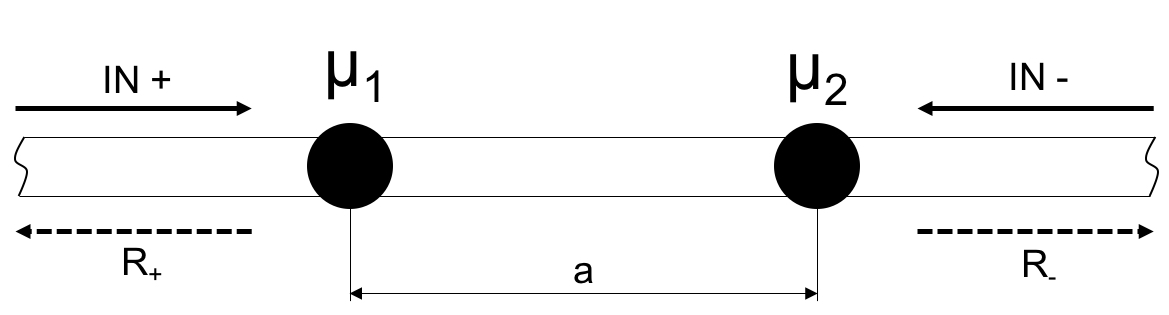}
	\caption{Schematic of the $N=2$ system.}
	\label{fig0}
\end{figure}

For $N=2$, let $x_1 = -\frac a2$, $x_2 = \frac a2$; a schematic of the system is shown in figure \ref{fig0}.
 The matrix ${\bf M}$ is
\beq{67-}
{\bf M} = \begin{pmatrix}
m_1 - g(0) & - g(a)
\\
- g(a) & m_2 - g(0)
\end{pmatrix}  
\eeq 
implying 
\bse{86-}
\bal{86}
R_{\pm} &= \frac {\ii}{2\det {\bf M}} 
\Big(   m_1  e^{\pm \ii k a} 
+       m_2   e^{\mp \ii k a} - {2g(0)\cos ka +2 g(a) }
\Big) ,
\\
T&= 1+ \frac {\ii}{2\det {\bf M}} 
   \big( m_1+m_2 - 2g(0) +2 g(a) \cos ka
\big).
\eal
\ese

The reflection coefficients can be written 
\bal{-23}
R_{\pm} =&  \frac {g^2(0)}{\det {\bf M}} 
\Big( e^{-ka} + \sin ka -  \cos ka -  (m_1+m_2)\cos ka  
\notag \\ & \pm \ii (m_2-m_1)\sin ka 
  \big)
\Big) .
\eal
This form shows that $|R_+| = |R_-|$ if $\mu_1$ and $\mu_2$ are real.  In that case there is no damping and the energy identity is satisfied: 
\beq{4=21}
|R_\pm|^2 + |T|^2 = 1 \ \ \text{for real }   \mu_1, \  \mu_2 .
\eeq
The reflection coefficients can vanish, for instance, if $ka =\pi$ and 
$m_1+m_2+ 1+e^{-\pi} = 0$, or if $m_1=m_2 = \frac 12 \big( 
\tan ka + e^{-ka}\sec ka -1 
\big)$.

When the separation $a\to 0$, Equation \eqref{-23}  yields $R_+ =R_-$ with the two attachments  acting as a single one with impedance $\mu = \mu_1+\mu_2$. 

\section{One way zero reflection: impedance conditions}\label{sec3}


We are interested in configurations in which one of the reflection coefficients vanishes but the other remains finite. {As shown above, the magnitudes of the reflection coefficients coincide for real-valued impedances.  Therefore, in order to have $|R_+| \ne |R_-|$ requires that at least one of the impedances $\mu_1,\ \mu_2$ is complex valued.}  We assume they are passive dampers, which means  that $\imag m_\alpha \le 0$ {for both $\alpha =1$ and $2$}.  
Assume $R_+ =0 $, then 
\beq{-11}
R_- =  \frac {(m_1-m_2)   }{\det {\bf M}} \sin ka
\eeq
subject to the constraint implied by $R_+ =0 $,
\beq{4-3}
e^{-ka} + \sin ka -  \cos ka -  (m_1+m_2)\cos ka +\ii (m_2-m_1)\sin ka =0 .
\eeq
Equivalently,
\beq{4-3a}
 m_+ \cot ka +\ii m_- =K
\eeq
where $m_+$, $m_-$ and {$K = 2\big( g(0) \cos ka - g(a)\big)/\sin ka$ are }   
\bal{3-99}
m_\pm &= m_1 \pm m_2 ,
\\
K &= 1 - \cot ka + e^{-ka} / \sin ka.
\eal
Viewing $m_+$ as the determining parameter, we have $R_+=0$ and 
{	\bse{92}
	\bal{4-}
	m_1 & = \frac {m_+}{2 } + \frac{\ii}2 (m_+ \cot ka - K) ,
	\\
	m_2 & = \frac {m_+}{2 } - \frac{\ii}2 (m_+ \cot ka - K),
	\\
	R_-&=  2\ii  \sin^2 ka\frac{ \big( [\frac 12  m_+  - g(0)]\cos ka  +g(a)\big)}
	{ \big(\frac 12  m_+  - g(0) +g(a)\cos ka \big)^2 
		}
	,\label{88}
	\\
T&= 1+\frac{  \ii \sin^2 ka}
	{ \frac 12  m_+  - g(0) +g(a)\cos ka  		} . 
	\label{8-8}
	\eal
	\ese }
One reason for considering $m_+$ to be the control parameter is that,
unlike $m_-$,  it must lie in the negative half of the complex plane, $\imag m_+\le 0$.  This does not, however,  guarantee that both $m_1$ and $m_2$ are in the same half plane.  Therefore the choice of $m_+$ must be restricted by the passivity requirements
  $\imag m_\alpha\le 0$,  $\alpha = 1,2$.  The case  $\imag m_+ = 0$ is of no interest, and we therefore concentrate on  $\imag m_+ < 0$.

Define
\beq{35}
m = m' + \ii m''
\eeq
then the reflection coefficient $R_+$ vanishes if 
\bse{2-44a}
\bal{-207a}
m_{+}' \cot ka - m_{-}'' &= K, 
\\
m_{+}'' \cot ka + m_{-}' &= 0 , \label{-207b}
\eal
\ese
where $ka \in (0, \pi)$. The non-zero  reflection coefficient $R_-$ of \eqref{-11}  can be rewritten as
\beq{-112}
R_- =  \frac { - m''_{+} \cos k a +\ii m''_{-} \sin k a   }{\det {\bf M}} ,
\eeq
indicating that damping  is essential in order to have $R_- \ne 0$.

\begin{figure}[h!]
	\includegraphics[width=0.78\linewidth]{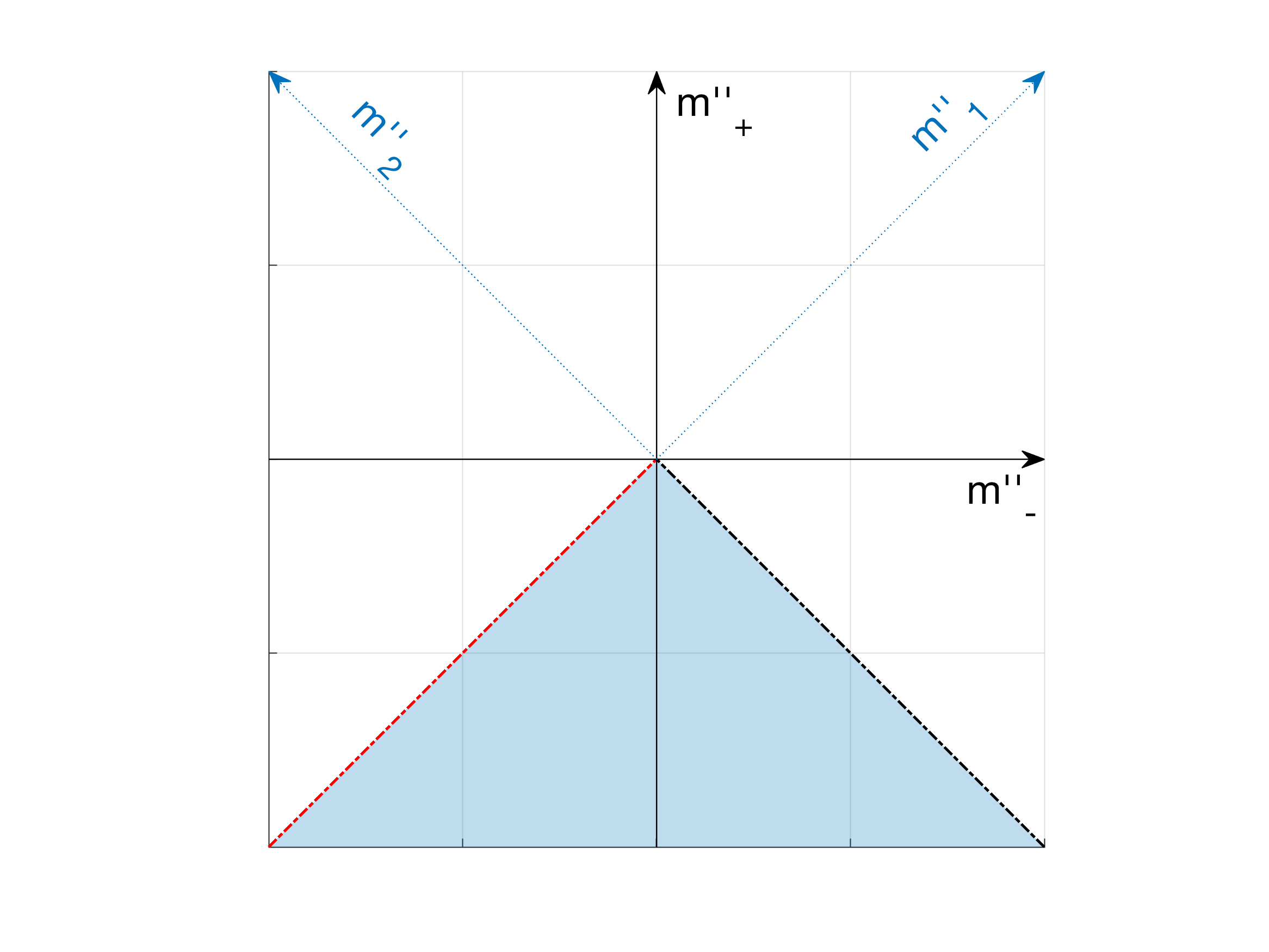}
	\caption{Design space for the two-scatterer case defined in terms of $( m''_+, m''_- )$ or $( m''_1, m''_2 )$. The shaded area marks all negative imaginary parts of $m_\alpha$ of the two scatterers, corresponding to passive damping properties.}
	\label{fig2}
\end{figure}

Of interest is therefore the part of the parameter space, $( m''_+, m''_- )$ or - equivalently - $( m''_1, m''_2 )$, where at least one of the scatterers displays passive damping properties, i.e. $m''_\alpha < 0$. Each point in this space uniquely defines the pair of scatterers in terms of their damping properties, but not their mass-stiffness properties or the spacing $a$ between them. Figure \ref{fig2} illustrates this space with the shaded area corresponding to the desired passive damping properties of the scatterers. It can be noted that $( m''_+, m''_- )$ and $( m''_1, m''_2 )$ coordinate systems are related by a $\pi/4$ rotation and $\sqrt{2}$ stretch through the transformation given in \eqref{3-99}. Passivity conditions, analogous to $m_{1}'' < 0$ and $m_{2}'' < 0$, are therefore given as $|m_{-}''| < - m_{+}''$.

\section{Examples of one way reflection} \label{sec4}


It was shown in section \ref{sec3} that damping is critical to obtain one way zero reflection. We will therefore consider the pair of scatterers $(\alpha, \beta)$ described by two complex normalized impedances $m_{\alpha,\beta} = m_{\alpha,\beta}' + \ii m_{\alpha,\beta}''$ with passive damping properties ($m_{\alpha,\beta}'' < 0$).

We start by investigating a setup composed of two scatterers where one is described by purely real $m_\alpha = m_\alpha'$, while the other by a complex normalized impedance $m_\beta = m_\beta' + \ii m_\beta''$, with $m_\beta'' < 0$. As the system is non-reciprocal, its response is non-symmetric with respect to the selection of $\alpha$ and $\beta$, namely for $( \alpha, \beta ) = (1,2)$ and $( \alpha, \beta ) = (2,1)$. We will therefore distinguish the two cases.

Next, we consider cases of two scatterers with passive damping properties, i.e. $m_{\alpha, \beta}'' < 0$. Starting with the case of the same negative normalized impedance $m_{\alpha}'' = m_{\beta}'' = m'' < 0$, we then generalize to $m_{\alpha}'' \neq m_{\beta}'' < 0$ and relate this general case to results obtained for other configurations of the scatterers.

\subsection{First impedance purely real: $m_1'' = 0$} \label{112=}
When $( \alpha, \beta ) = (1,2)$, we have   $m_1 = m_1'$, $(m_1'' = 0)$  and $m_2 = m_2' + \ii m_2''$, with $m_2'' < 0$, i.e. we investigate scatterer configurations along an edge of the passive zone (see the black dash-dot line in figure \ref{fig2}). Then, $m_{+}'' = m_2''$ and $m_{-}'' = - m_2''$ and the relation between $m_{+}'$ and $m_{-}'$ follows from \eqref{2-44a} as
\beq{1001}
\big(   m_{+}' \cot ka -K\big) \cot ka - m_{-}' = 0, \ \ \ \ \ \ \ \ m_2'' = m_{+}'' = - m_{-}'' < 0.
\eeq
From equations \eqref{2-44a} and \eqref{1001} we also have that
\beq{1002}
m_{-}' \gtrless 0, \ \ m_{+}' \cot ka \gtrless K, \ \ \text{for} \ \  
\left\{\begin{matrix}
k \in (0, \pi/2),
\\
k \in (\pi/2, \pi).
\end{matrix}  \right.
\eeq
Equation \eqref{1001} with constraints in \eqref{1002} define a surface in $(m_+', m_-')$ space that  is illustrated in figure \ref{fig3}. Relation \eqref{1001} can be mapped to $(m_{1}',m_{2}')$ coordinates through \eqref{3-99} to find corresponding real parts $m_1'$ and $m_2'$ of the complex impedances. Two particular configurations can be then obtained by cutting the design space from figure \ref{fig3} with the $m_{1}' = 0$ or $m_{2}' = 0$ planes resulting in one of the impedances being purely real and the other purely imaginary.

\subsubsection{Second impedance purely imaginary.} \label{sp1}  
\begin{figure}[h!]
	\includegraphics[width=0.78\linewidth]{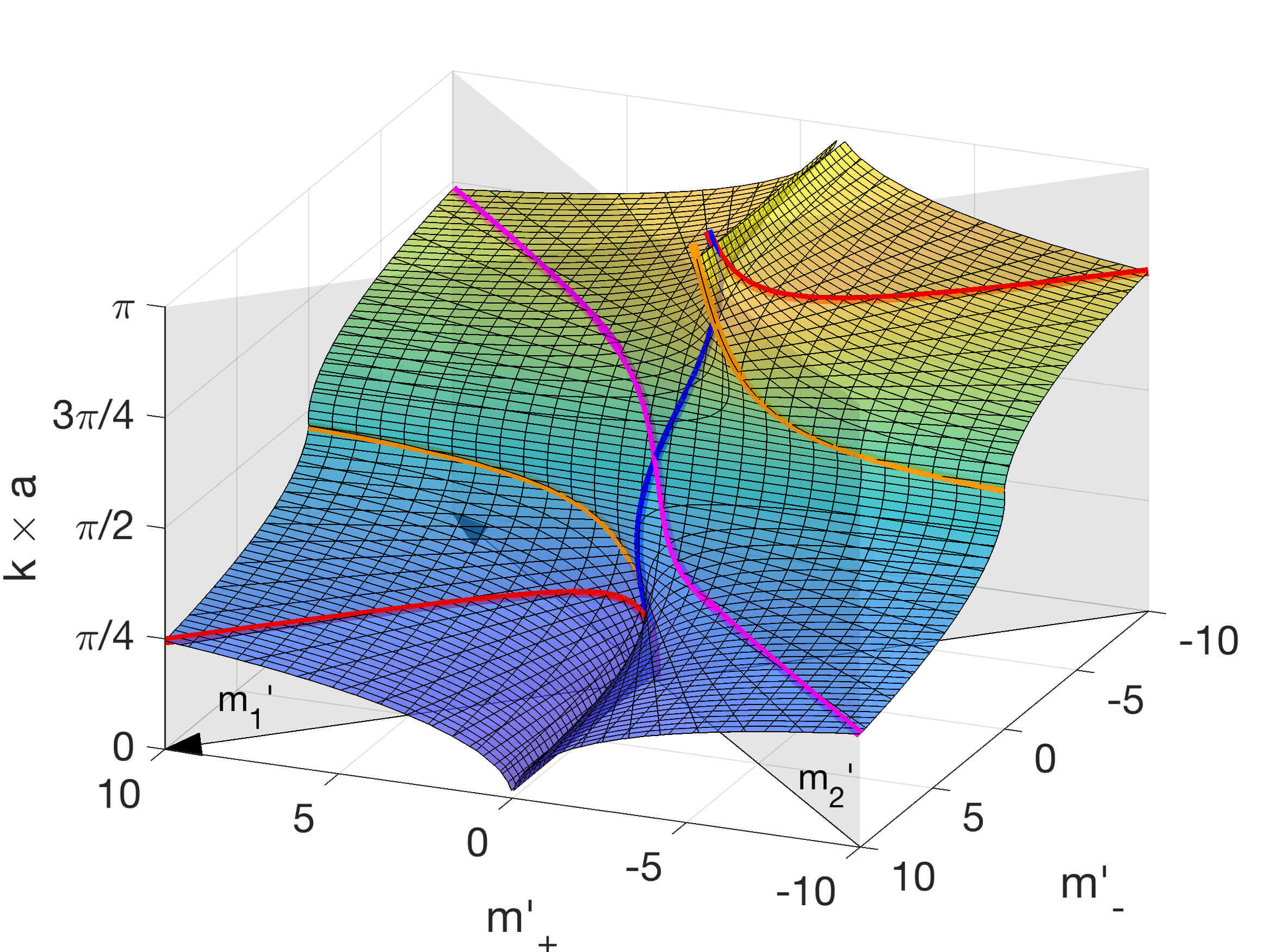}
	\caption{Design space, $|m_-''| < - m_+''$, corresponding to negative imaginary parts of $m_1$ and $m_2$ mapped through equations \eqref{2-44a} (see figure \ref{fig2}). The color lines indicate  special configurations of scatterers described in the text.  Solutions for the red, magenta, blue and orange curves are given in figures \ref{fig1}, \ref{fig4}, \ref{fig5}  and \ref{fig18}, respectively.}
	\label{fig3}
\end{figure}

\begin{figure}[h!]
	\includegraphics[width=0.78\linewidth]{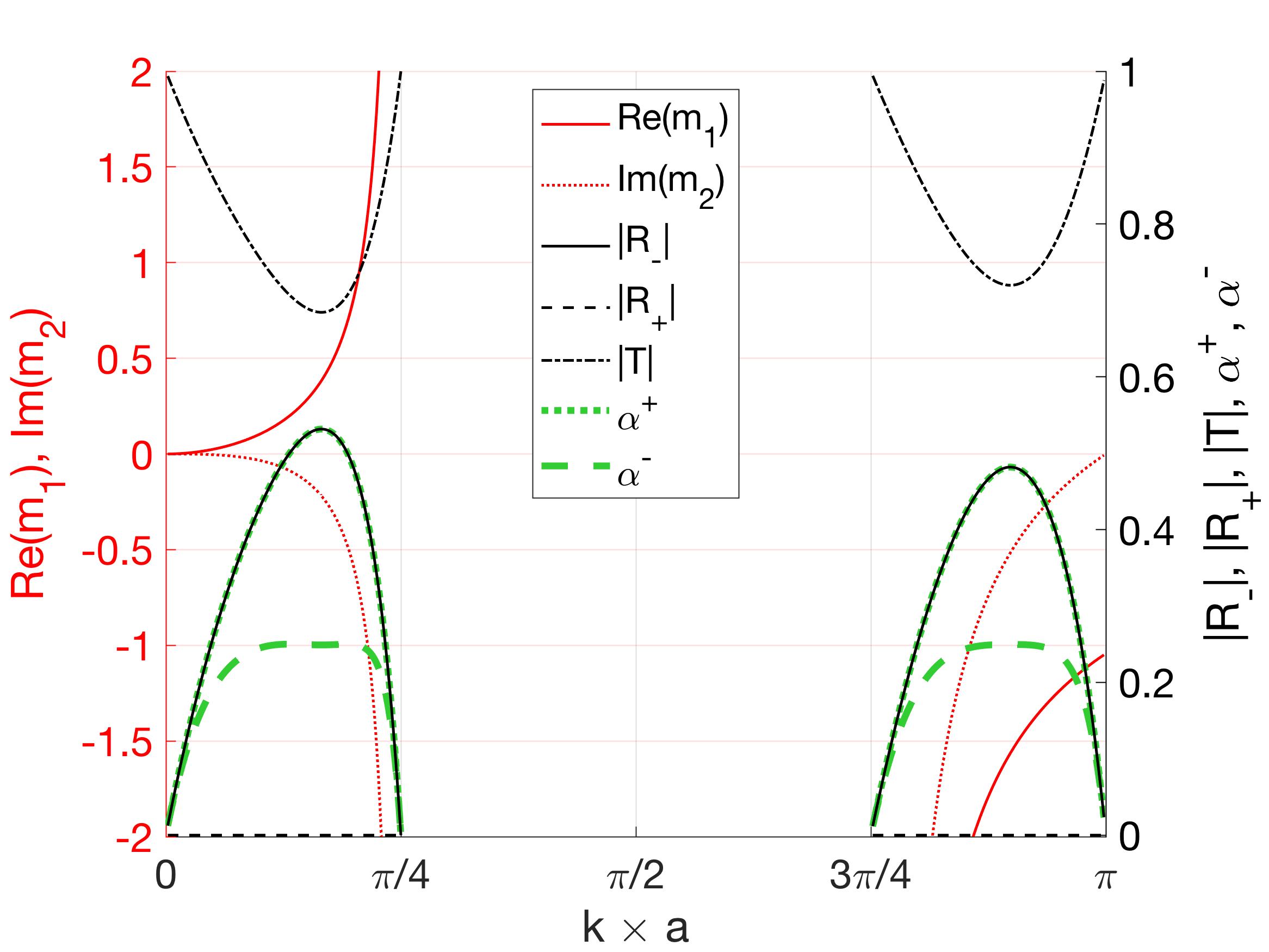}
	\caption{The normalized impedances for zero one-way reflection from equation \eqref{-78}. In this case $\mu_1  $ is real and $\mu_2 $ is positive imaginary and hence passive. }
	\label{fig1}
\end{figure}

Continuing with  $(\alpha, \beta) = (1,2)$ (impedance $\mu_1$ is purely real), but specializing to the case of  $\mu_2$  purely imaginary $(m_1'' = m_2'=0, \, m_-= \overline{ m}_+)$, corresponds to a particular solution of the family of solutions illustrated in figure \ref{fig3} by red lines. Then \eqref{2-44a} implies that 
\beq{-78} 
m_1' = \frac{K \cot ka}{\cot^2 ka - 1}, 
\ \ \
m_2'' = \frac{- K}{\cot^2 ka - 1}.
\eeq
Figure \ref{fig1} shows the required values of $m_1'$ and $m_2''$ and the corresponding reflection and transmission coefficients, $|R_{+}|$, $|R_{-}|$ and $|T|$. The impedance $\mu_2$ for this model is passive for $ka \in (0,\frac{\pi}4)$ and $ka \in (\frac{3\pi}4,\pi)$. Note that the intersection of the surface given by \eqref{1001} with $m_1' = 0$ reduces to a point at $k   a = 0$.

\subsection{Second impedance purely real: $m_2'' = 0$}  \label{113=}  
If $(\alpha,\beta) = (2,1)$, then $m_2'' = 0$ ($m_2 = m_2'$) and $m_1 = m_1' + \ii m_1''$ with $m_1'' < 0$ corresponding to the red dash-dot line in figure \ref{fig2}. In this case $m_{+}'' = m_1''$ and $m_{-}'' = m_1''$ and the relation between $m_{+}'$ and $m_{-}'$ yields
\beq{1003}
 \big(   m_{+}' \cot ka -K\big) \cot ka + m_{-}' = 0, \ \ \ \ \ \ \ \ m_1'' = m_{+}'' = m_{-}'' < 0,
\eeq
while equations \eqref{2-44a} and \eqref{1003} imply the constraints
\beq{1004}
m_{-}' \gtrless 0, \ \ m_{+}' \cot ka \lessgtr K, \ \ \text{for} \ \  
\left\{\begin{matrix}
k \in (0, \pi/2),
\\
k \in (\pi/2, \pi).
\end{matrix}  \right.
\eeq
The design space given by \eqref{1003} is shown in figure \ref{fig3}. As before, two particular configurations can be seen by cutting the design space from figure \ref{fig3} by the $m_{1}' = 0$ or $m_{2}' = 0$ planes, subject to the constraints  \eqref{1004}.

\subsubsection{First impedance purely imaginary.} \label{sp2}

\begin{figure}[h!]
	\includegraphics[width=0.78\linewidth]{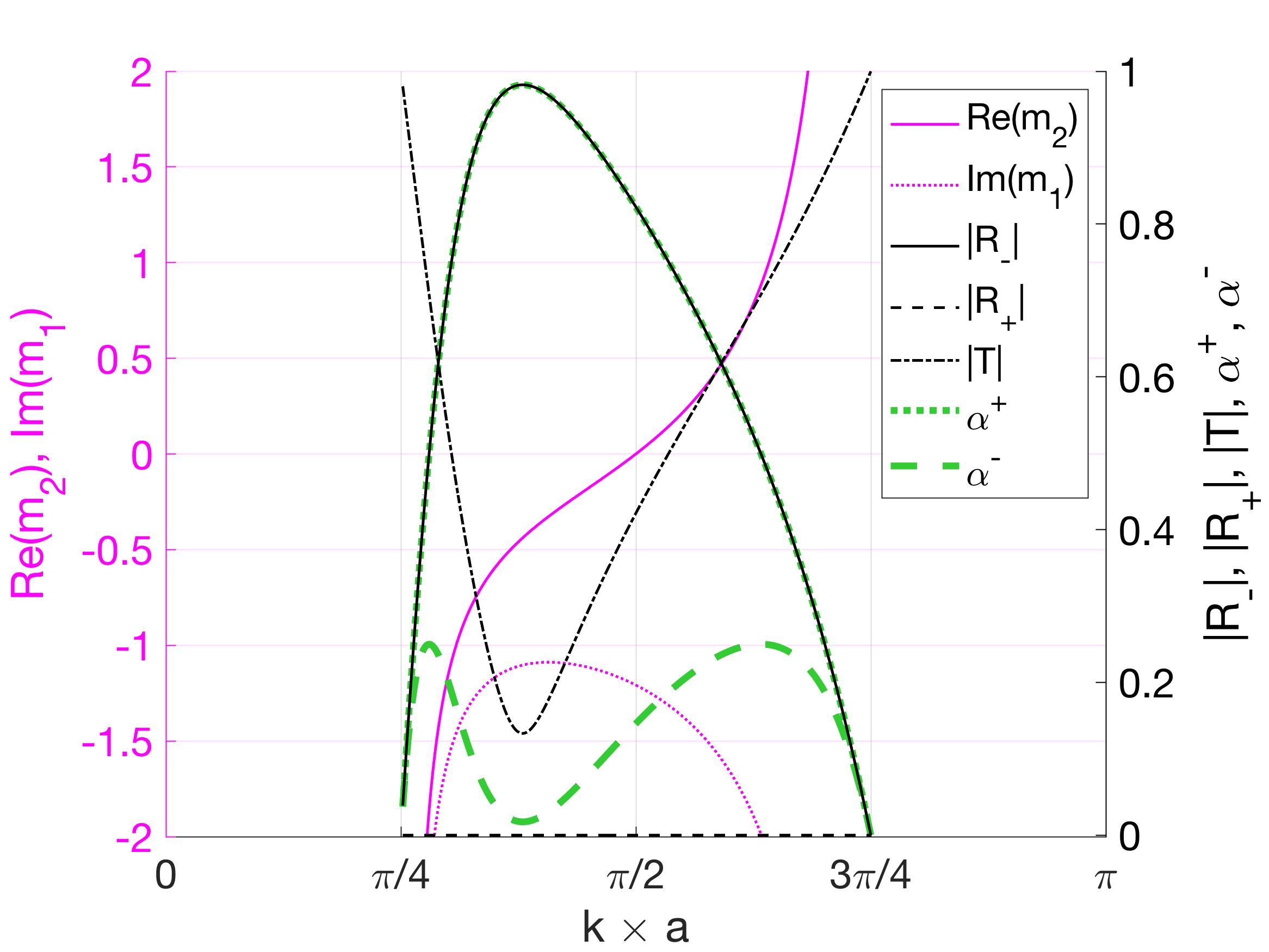}
	\caption{The normalized impedances for zero one-way reflection from equation \eqref{1005}. In this case $\mu_2  $ is real and $\mu_1 $ is positive imaginary and hence passive. }
	\label{fig4}
\end{figure}

When mapping \eqref{1003} on the $m_2'$ plane, i.e. taking $m_1' = 0$, we have $m_1 = \ii m_1''$, $m_1'' < 0$ and $m_2 = m_2'$.  The first impedance purely negative imaginary while the second is purely real $(m_-= -\overline{ m}_+)$, a situation opposite to that considered previously, with  
\beq{1005}
m_1'' = \frac{K}{\cot^2 ka - 1}, 
\ \ \
m_2' = \frac{ K \cot ka}{\cot^2 ka - 1}.
\eeq
Figure \ref{fig4} shows the required values of $m_1''$ and $m_2'$ and the corresponding reflection and transmission coefficients, $|R_{+}|$, $|R_{-}|$ and $|T|$.

\subsubsection{Second impedance infinite (pinned point).} \label{sp3} 

Due to asymmetry of the design space, another interesting scatterer configuration can be found by mapping \eqref{1003} onto the $m_1'$ plane. Then, $m_2' = 0$, making the second normalized impedance  vanish and resulting in $\mu_2 = \infty$, i.e. a pinned point. Note that from the definition of the impedance and the governing equation \eqref{-214} it follows that energy can be transferred across a pinned point through rotations of the beam cross sections proportional to $\frac{d w}{d x}$ even if $w= 0$. {This particular feature distinguishes the flexural wave problem from the acoustic one.} In this case the other normalized impedance, $m_1 = m_1' + \ii m_1''$, is complex with  
\beq{1006}
m_1' = K \sin ka \cos ka, 
\ \ \
m_1'' = {-K}{\sin^2 ka }.
\eeq

\begin{figure}[h!]
	\includegraphics[width=0.78\linewidth]{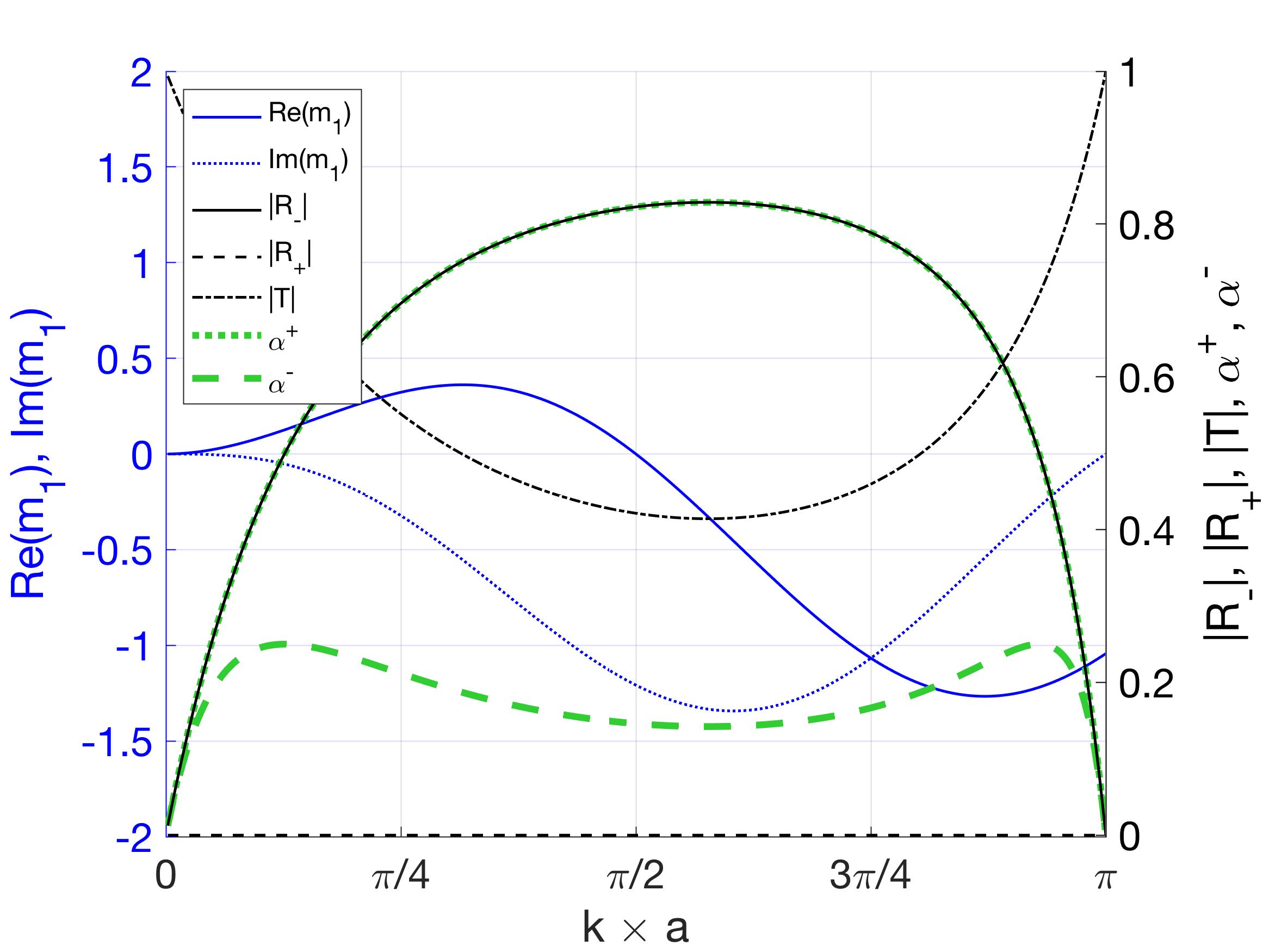}
	\caption{The normalized impedances for zero one-way reflection from equation \eqref{1006}, corresponding to the blue curve in figure \ref{fig3}. In this case $m_2 = 0$ ($\mu_2 = \infty$)  indicating  a pinned point,  and $m_1 = m_1' + \ii m_1''$ is complex with $m_1'' < 0$ and hence passive. }
	\label{fig5}
\end{figure}

Figure \ref{fig5} shows the real and imaginary parts of the complex normalized impedance for the first scatterer. Interestingly, with the second point pinned, $m_2 = 0$, it is possible to obtain one-way reflection over wide wavenumber (or frequency) band.

\begin{table*}[t]
  \centering
\begin{tabular}{|c|c|c|c|c|c|c|} 
\hline  
Passive $ka$  &$m_1'$ &  $m_1''$ & $m_2'$  & $m_2''$   &  Figure & Sec. 
\\
\hline   \hline
$(0,\frac{\pi}4), (\frac 34 {\pi}, \pi)$ & $\frac K2 \tan 2ka$&  $0$ & $0$  & $-\frac K2 \tan 2ka  \tan ka$ &  \ref{fig1} & \ref{sp1}
\\  \hline
$(\frac{\pi}4 , \frac 34 \pi)$ & $0$ & $\frac K2 \tan 2ka  \tan ka $ & $\frac K2 \tan 2ka$ & $0$ &  \ref{fig4}  & \ref{sp2}
\\  \hline
$(0,{\pi})$ &$\frac K2 \sin 2ka$ & {$- K \sin^2 ka$}  &  $0$  &  $0$ &   \ref{fig5}  & \ref{sp3}
\\  \hline
$(0,{\pi})$ &$\frac K2 \tan ka - m''\cot ka$ & $m''$  &  $\frac K2 \tan ka + m''\cot ka$  &  $m''$ &   \ref{fig6},\ref{m1m2},\ref{rtcase3}  & \ref{sec02}
\\  \hline
$(0,{\pi})$ &$K \tan ka$ & $- \frac K2 \tan^2 ka$  &  $0$  &  $- \frac K2 \tan^2 ka$ &   \ref{fig7}  & \ref{sp4}
\\  \hline
$(0,{\pi})$ &$\frac K2 \tan ka \pm \Delta_{-}^{(2)}$ & $\mp \Delta^{(0)}_+ \tan ka $  &  $\frac K2 \tan ka \mp \Delta_{+}^{(2)}$  &  $\mp  \Delta^{(0)}_- \tan ka$ &   \ref{fig18}, \ref{fig18a}  & \ref{pert}
\\  \hline
${ka \ll 1 \atop ka \approx \pi}$ &$\frac K2 \tan ka - m''\cot ka$ & $m'' \ll 1$  &  $\frac K2 \tan ka + m''\cot ka$  &  $m'' \ll 1$ & \ref{asymptotics}  & ${\ref{sec51} \atop \ref{sec52}}$
\\  \hline
$ \approx (\pi, 2\pi)$ & $m_2'+\frac 12 \sin 2ka$ & $- m_2'' -\sin^2 ka $ & $-\frac 12 +\frac 12 e^{-ka}\cos ka$ & $\frac 12 e^{-ka}\sin ka$ &     & \ref{sec6}
\\  
\hline \hline
\end{tabular}
\caption{A summary of the special cases corresponding to the red, blue, magenta and orange curves in Figure \ref{fig3}, considered in this work.  Note $m_1'' \le 0, m_2'' \le 0$, ensuring passive configurations and 
$\Delta^{(n)}_{\pm} = \frac{\delta}{2} \big( 1 \pm \frac 1c{\tan^{n} ka} \big)$ for $n=0,2$. }
\label{table1}
\end{table*}
 

\subsubsection{Second impedance infinite (pinned point) with a pure damper.}  \label{m1im2izero}
Note the exchanged positions (but the same values) of the real parts of normalized impedances shown in rows 2 and 3 of table \ref{table1}. A particular selection of $k \times a = \pi / 2$ results in the real parts of scatterers in rows 2 and 3 - $m'_2$ and $m'_1$, respectively - equal to zero. Therefore, for this specific configuration we have the second scatterer pinned ($\mu_2 = \infty$) while the first is purely negative imaginary (passive damper). The imaginary parts of the normalized impedances $m''_1$ from \eqref{1005} and \eqref{1006} are both $m''_1 = - K = -(1 + e^{- \pi / 2}) < 0$.

\subsection{Impedances with the same passive damping properties:
$m_1'' = m_2''$}\label{sec02}

\begin{figure}[h!]
	\includegraphics[width=0.78\linewidth]{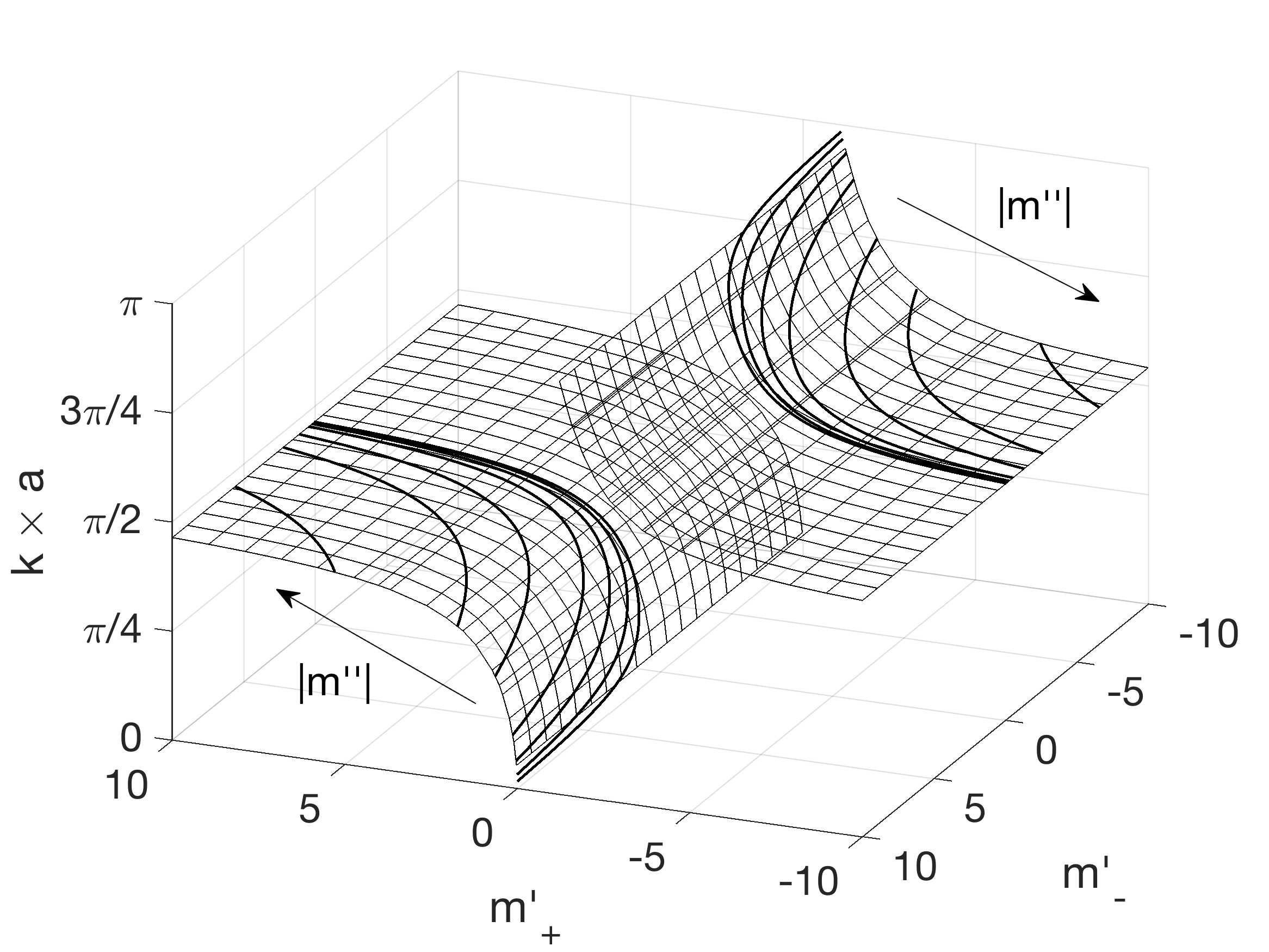}
	\caption{Examples of normalized impedance curves for various selected values of $m'' < 0$ for $m_+'' < 0$ and $m_-'' = 0$, based on Equation \eqref{1900}.}
	\label{fig6}
\end{figure}

Selecting passive damping properties other than discussed above, results in configurations with imaginary parts of normalized impedances being both nonzero. A particular choice is $m_1'' = m_2'' < 0$ or, equivalently $m_-'' = 0$ and $m_+'' < 0$. With $m_1'' = m_2'' = m'' < 0$ equations \eqref{2-44a} lead to
\beq{3-57}
m_+ = K \tan ka + \ii 2 m'' . 
\eeq
Therefore from \eqref{92}, the two impedances are 
\beq{1900}
\begin{aligned}
m_1 &= \frac 12 K \tan ka - m'' \cot ka +\ii m'' , 
\\
m_2 &= \frac 12 K \tan ka + m'' \cot ka +\ii m'' ,
\end{aligned}
\eeq
implying $|m_1'| > |m_2'|$. It is therefore not possible to obtain one-way reflection with a pair of scatterers with the same negative imaginary part of the complex impedance  having the same or opposite real parts (i.e. mass and stiffness properties). 
 Equation \eqref{1900} implies 
\beq{1007a}
(m_1')^2 - (m_2')^2 = -2 m'' K 
\ \ \Leftrightarrow \ \  
m_+' m_-' = -2 m'' K
\eeq
 which defines a hyperbola in $(m_1',m_2')$  (or $(m_+',m_-')$) space for each selected value of $m''$, as shown in figure \ref{fig6}.

\begin{figure}[h!]
	\centering
	\begin{subfigure}{}
		\includegraphics[width=0.48\linewidth]{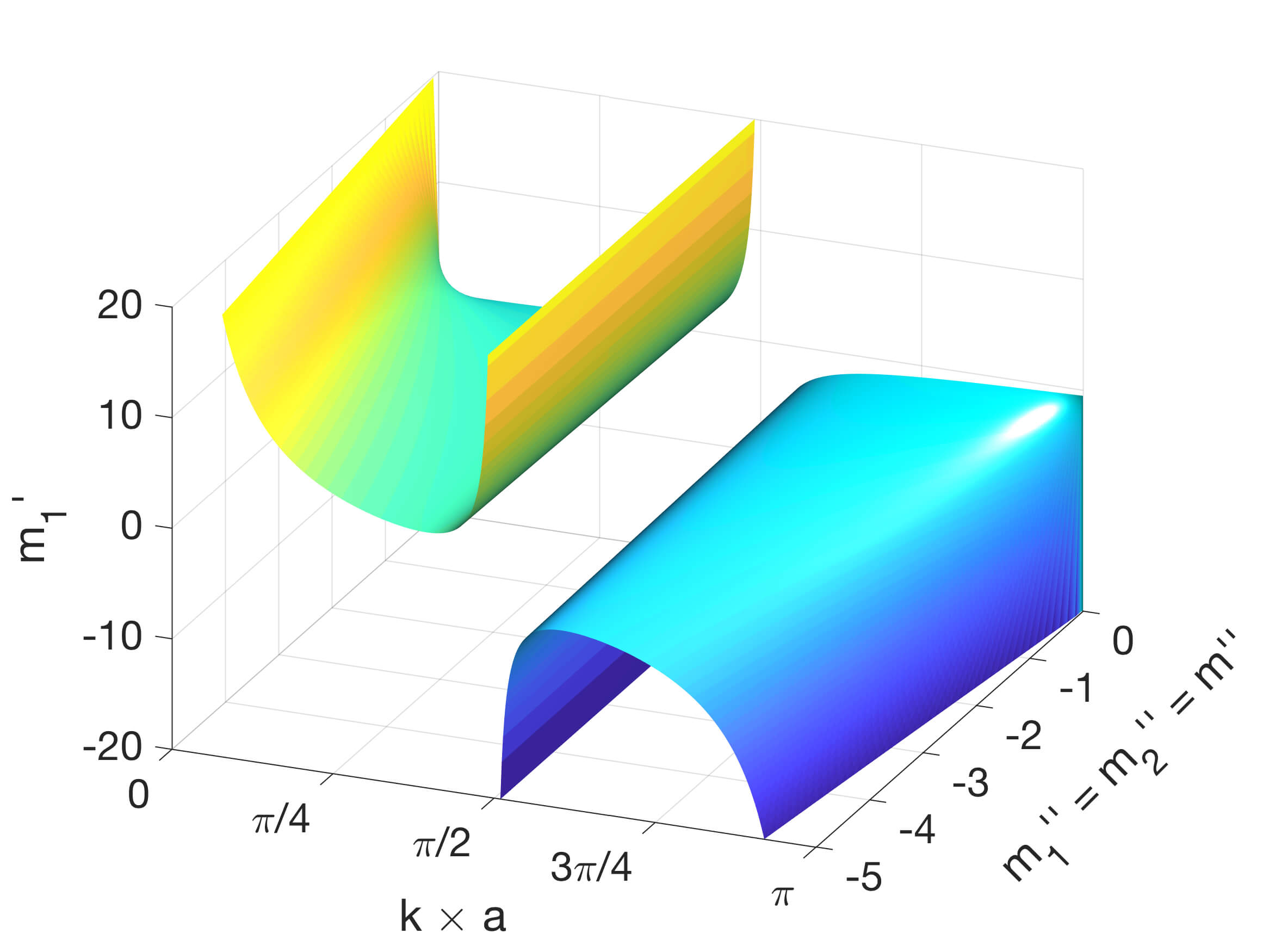}
	\end{subfigure}
	\begin{subfigure}{}
		\includegraphics[width=0.48\linewidth]{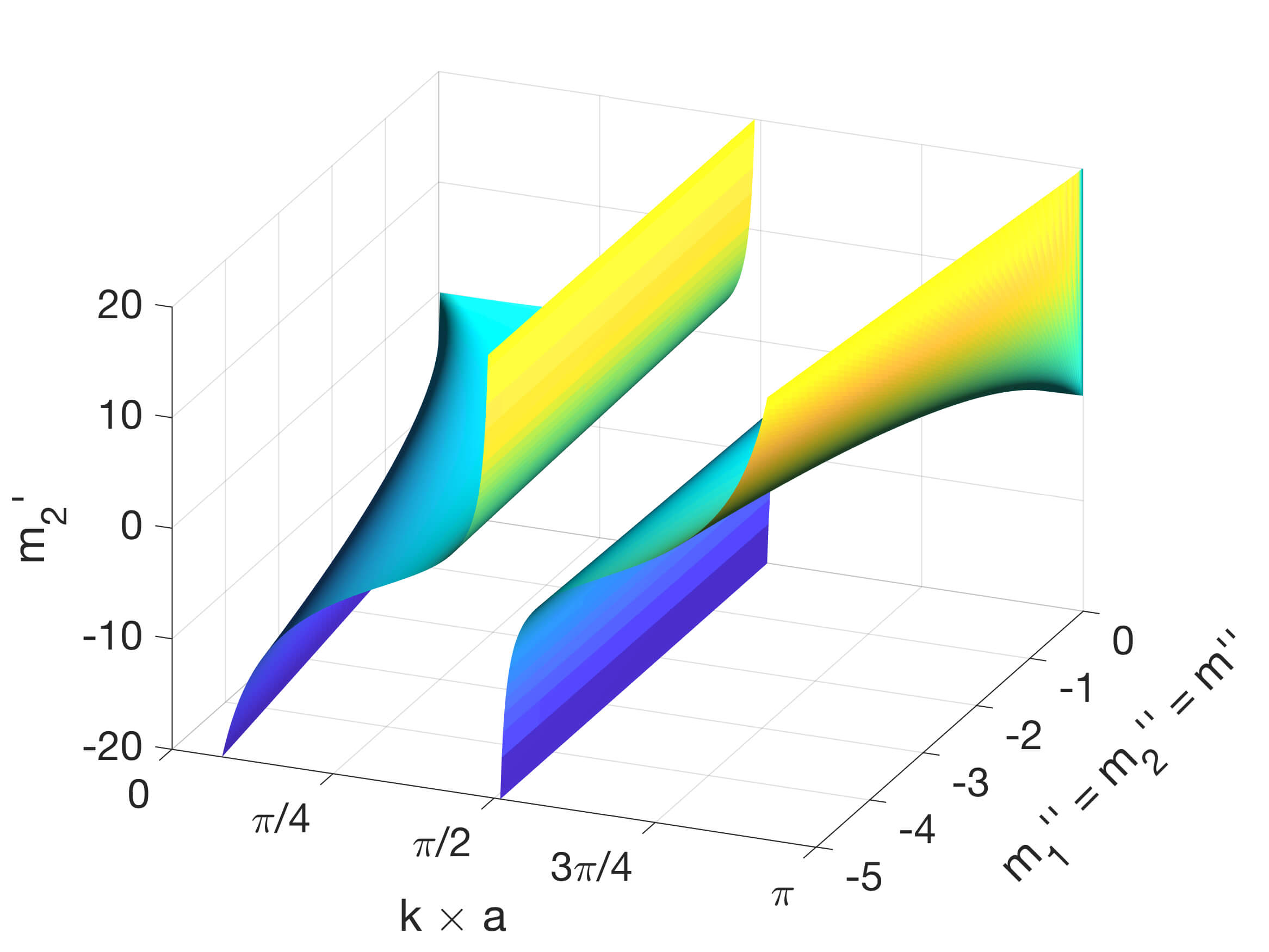}
	\end{subfigure}
	\caption{Values of $m_1'$ and $m_2'$ required for one-way zero reflection when $m_1'' = m_2'' = m'' < 0$, computed from equation \eqref{1900}.}
	\label{m1m2}
\end{figure}

\begin{figure}[h!]
	\centering
	\begin{subfigure}{}
		\includegraphics[width=0.48\linewidth]{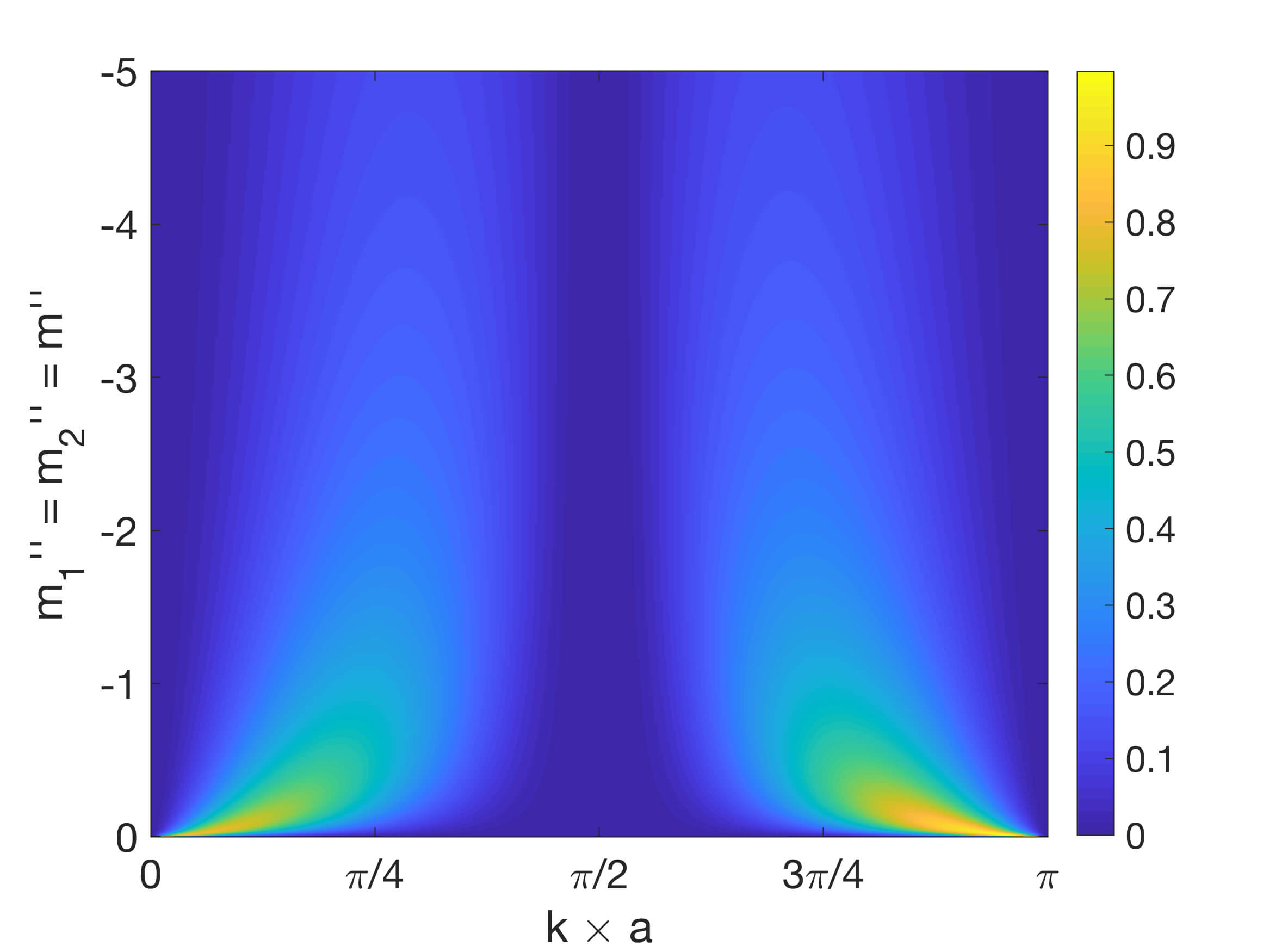}
	\end{subfigure}
	\begin{subfigure}{}
		\includegraphics[width=0.48\linewidth]{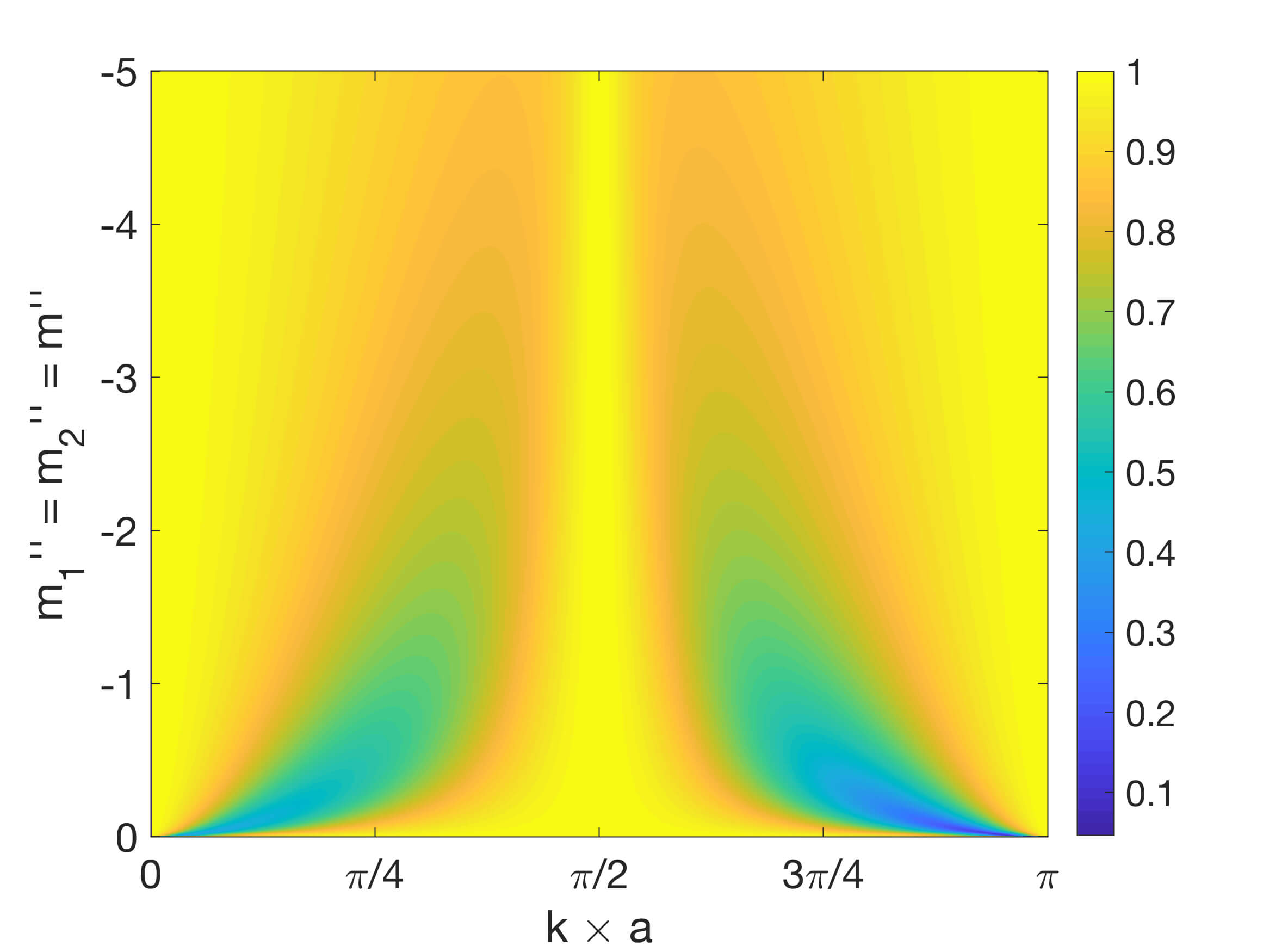}
	\end{subfigure}
	\caption{Reflection $|R_{-}|$, (a), and transmission $|T|$, (b), for $m_1'' = m_2'' = m'' < 0$; $|R_{+}| = 0$. Note the large reflectivity for small $m'' $ and for $ka\approx 0, \, \pi$. This phenomenon is examined in Section \ref{sec5}.}
	\label{rtcase3}
\end{figure}

Values of $m_1'$ and $m_2'$ required for the one-way zero reflection with $|R_{+}| = 0$ are shown in figure \ref{m1m2}. The corresponding reflection $|R_{-}|$ and transmission $|T|$ coefficients are shown in figure \ref{rtcase3}. Note the possibility of obtaining narrow-band highly directional properties of the system when $m'' \rightarrow 0$ and $k \times a \rightarrow 0$ or $k \times a \rightarrow \pi$. Those special cases are discussed in detail later.

\subsubsection{Second impedance purely imaginary.}  \label{sp4}
\begin{figure}[h!]
	\includegraphics[width=0.78\linewidth]{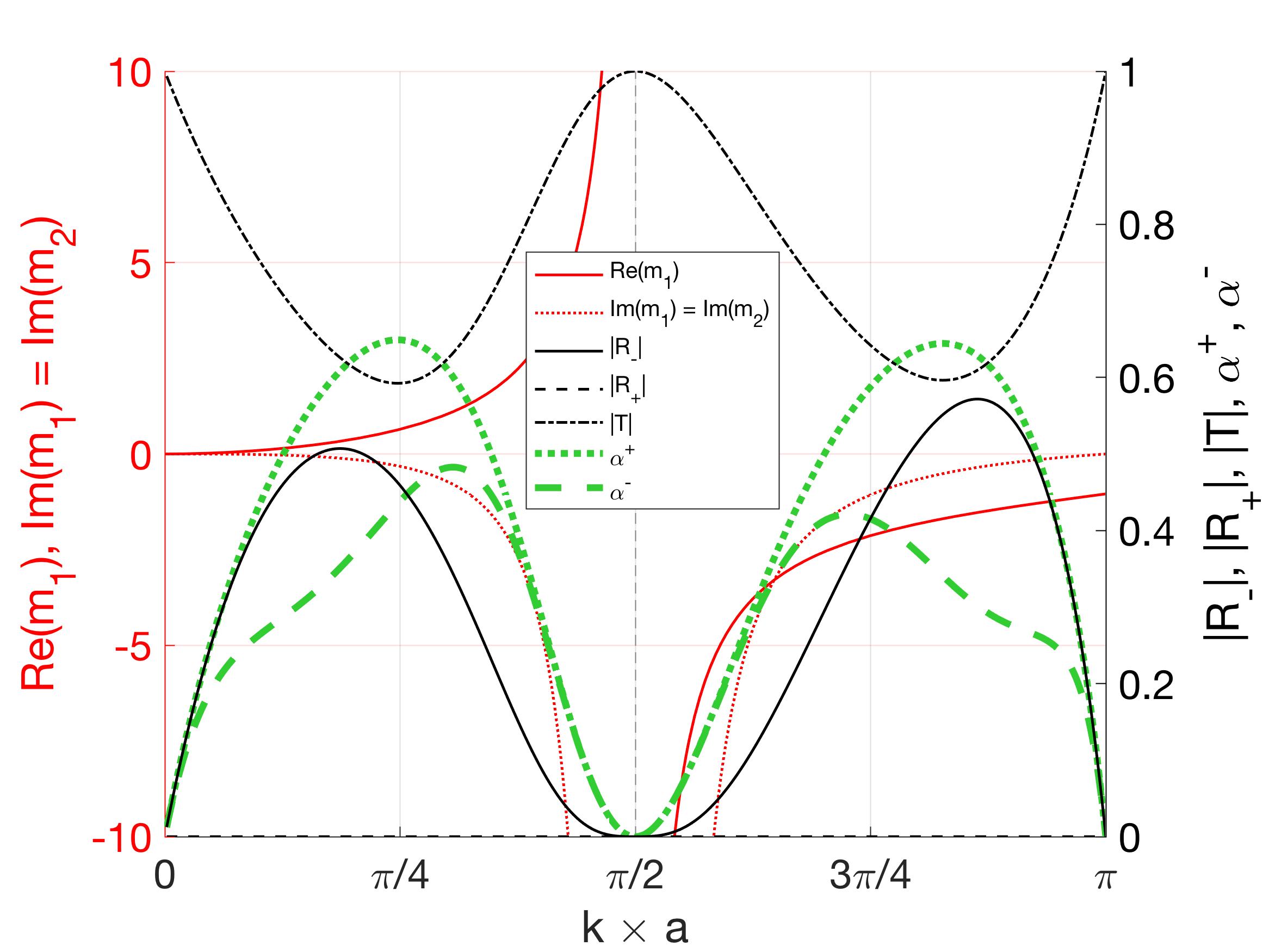}
	\caption{The normalized impedances for zero one-way reflection from equations \eqref{1901}. In this case the two scatterers have the same damping properties, $m_1'' = m_2'' = m''$ and $m_2' = 0$.}
	\label{fig7}
\end{figure}

A special case of scatterer configuration can be obtained for $m_2' = 0$. Then, with $m_1 = m_1' + \ii m''$ and $m_2 = \ii m''$ we have two scatterers with the same damping properties. Equation \eqref{1007a} reduces to
\beq{1007b}
(m_1')^2 = -2 m'' K
\eeq
and uniquely defines the relation between $m''$ and $m_1'$. Equations \eqref{1900} can be then reduced to
\beq{1901}
m_1' = K \tan ka, 
\ \ \
m'' = - \frac 12 K \tan^2 ka .
\eeq
Possible choices of $m_1'$ as a function of $m''$ and $k  a$ are shown in figure \ref{fig7}.

\subsection{Almost equal  impedances.} \label{pert}

We now analyze the case where the differences between the real and imaginary parts of the normalized impedances are small. In particular we are interested in the cut of the design space for $m_{-}' \approx 0$, as shown by the orange lines in figure \ref{fig3}. We therefore set $m_{-}' = \pm \delta$ 
where $\delta$ is a small number allowing for shifting the cutting plane in the design space; the upper sign is taken for $k \times a \in (0, \pi / 2)$ and lower for $k \times a \in (\pi / 2, \pi)$, and are required for passive damping, 
i.e.\ $m_{+}'' < 0$ where $m_{+}'' $ follows from  
  \eqref{-207b}.
At the same time  $m_{-}''$ must satisfy $|m_{-}''| < - m_{+}''$, which is  imposed   by setting
\beq{w3}
m_{-}'' = \frac{m_{+}''}{c} = \mp \frac{\delta}{c} \tan ka, \ \ \ \ c \in (- \infty, -1) \cup (1, \infty)
\eeq
with $c \in (- \infty, -1)$ for $m_{-}'' > 0$ and $c \in (1, \infty)$ for $m_{-}'' < 0$, and $m_{+}'$ can then be obtained by using \eqref{-207a}.
In summary,
\beq{4=7}
\begin{aligned}
m_+ & = \big[ K \mp \delta \big( \frac{\tan ka}{c}  +\ii\big)  \big] \tan ka ,
\\
m_- & = \pm \delta  \big( 1 - \ii \frac{\tan ka}{c} \big),
\end{aligned}
 \ \  \ \text{for }\ 
\left\{\begin{matrix}
k \in (0, \pi/2),
\\
k \in (\pi/2, \pi).
\end{matrix}  \right.
\eeq

From $m_{-}' = \pm \delta$ and \eqref{w3} it can be seen that for small $\delta$ we have both real and imaginary parts of the normalized impedances nearly the same, regardless of the choice of $c$. Before discussing the case of $k \times a \rightarrow 0$ and $k \times a \rightarrow \pi$ in the next section, we present $m_1$ and $m_2$ as functions of $k \times a$, required for the zero one-way reflection.

\begin{figure}[h!]
	\centering
	\begin{subfigure}{}
		\includegraphics[width=0.48\linewidth]{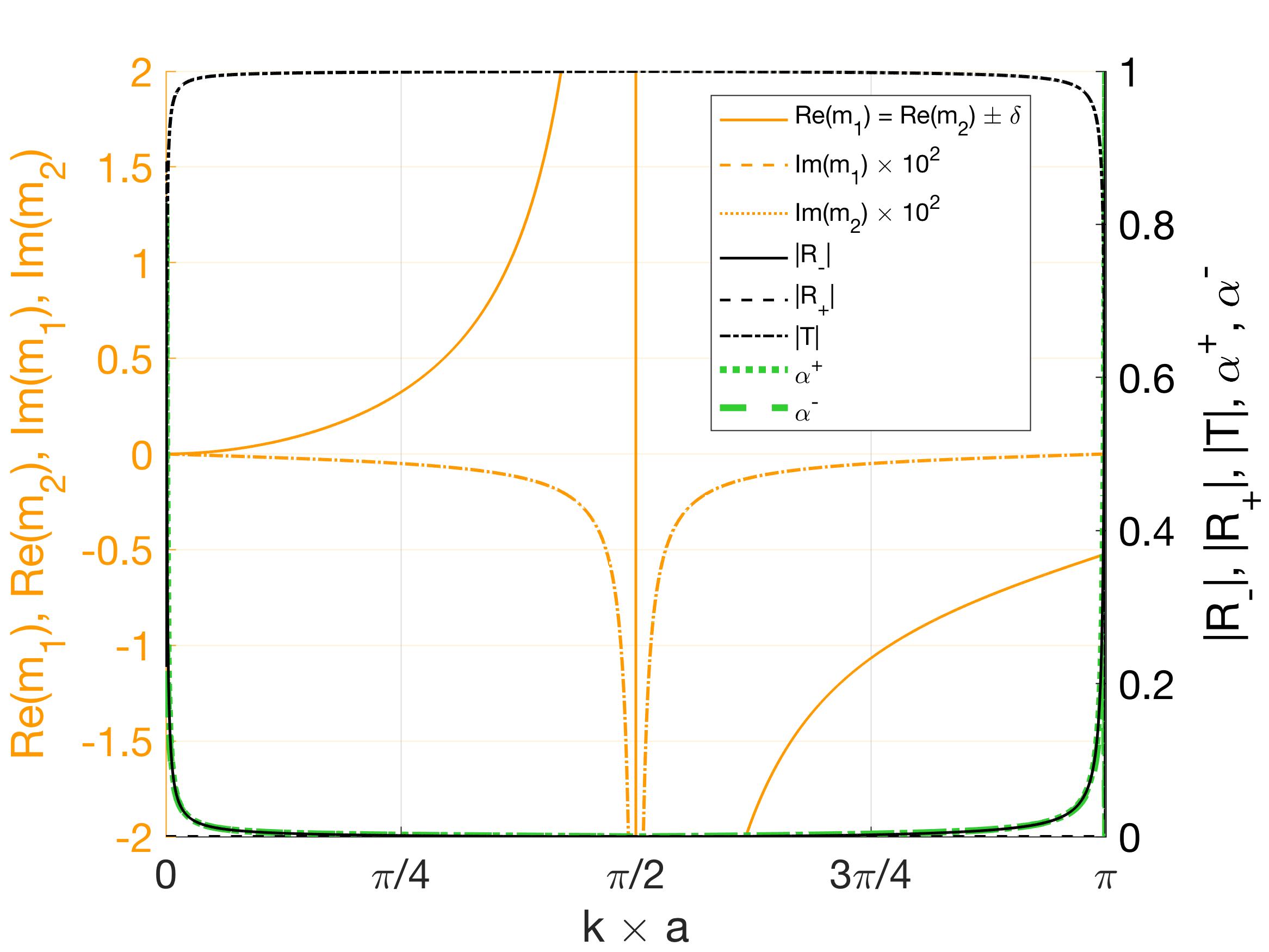}
	\end{subfigure}
	\begin{subfigure}{}
		\includegraphics[width=0.48\linewidth]{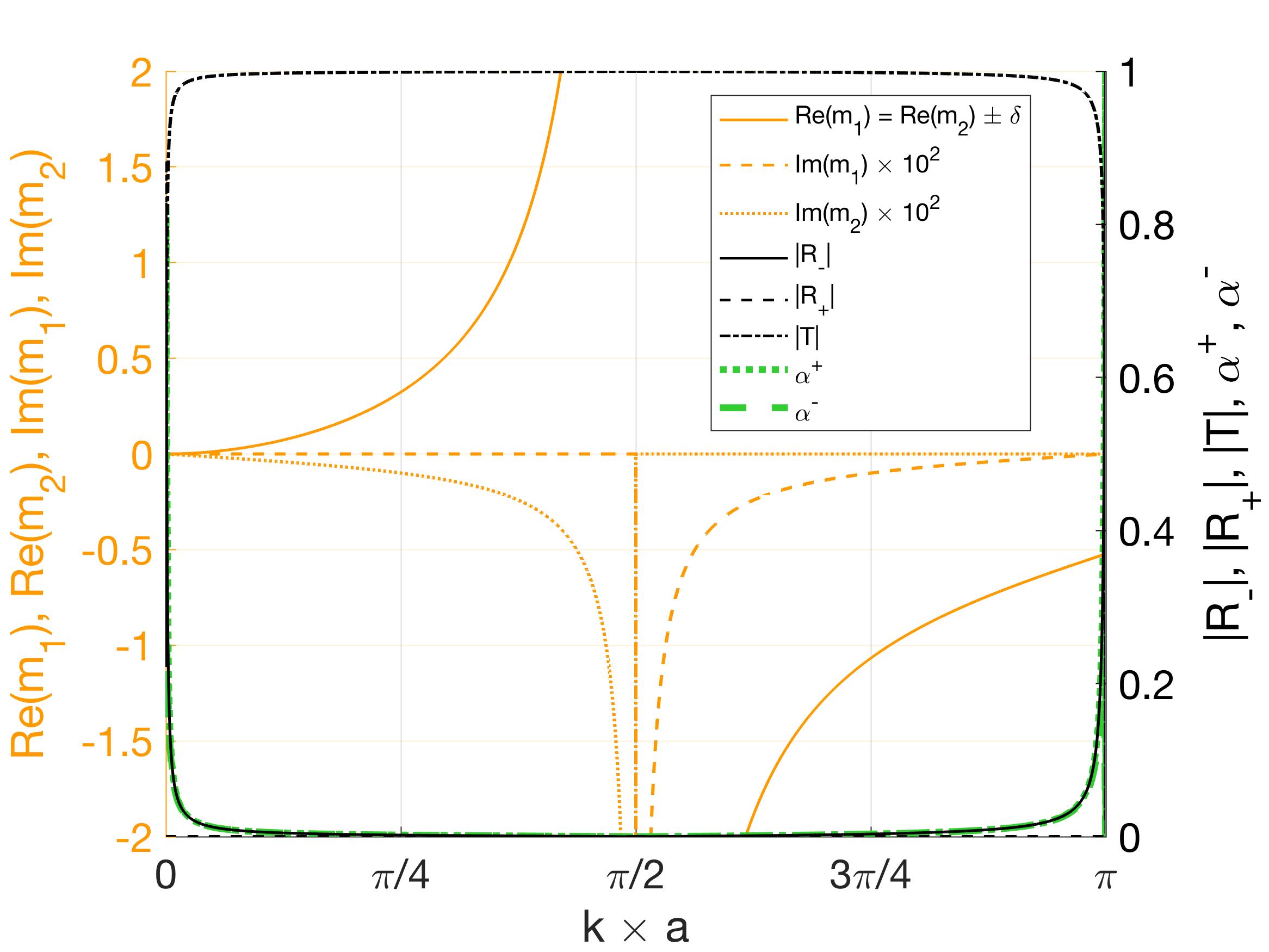}
	\end{subfigure}
	\caption{The normalized complex impedances for zero one-way reflection  with $\delta = 1 \times 10^{-3}$ and $c = 1000$ (a), and $c = 1$ (b). In this case the scatterers share close real and imaginary parts of the complex impedances (a), and close values of the real parts with one of the imaginary parts close to zero (b).}
	\label{fig18}
\end{figure}

Figure \ref{fig18} shows the real, $m_1'$, $m_2'$, and imaginary, $m_1''$, $m_2''$, parts of the normalized complex impedances for two selected combinations of $(\delta,c)$, namely $(1 \times 10^{-3}, 1 \times 10^3)$ (figure \ref{fig18}a) and $(1 \times 10^{-3}, 1)$ (figure \ref{fig18}b), along with reflection and transmission coefficients. For large $c$ (figure \ref{fig18}a) both real and imaginary parts assume nearly the same values over wide $k \times a$ range. For small $c$ (figure \ref{fig18}b), scatterers configurations analogous to those presented in sections \ref{112=} and \ref{113=}, are obtained with the difference in assumption on the real parts being nonzero and having close values. Note all the configurations presented in figure \ref{fig18} are passive.

\begin{figure}[h!]
	\includegraphics[width=0.98\linewidth]{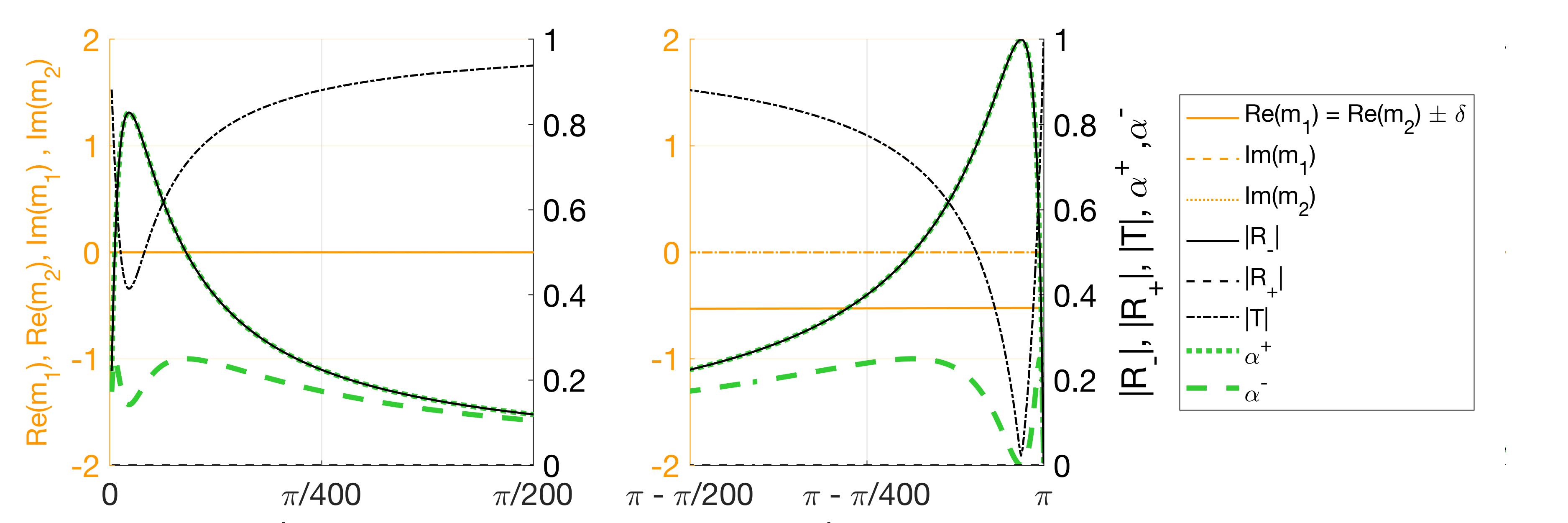}
	\caption{A close-up view of figure \ref{fig18} (a), ($\delta = 1 \times 10^{-3}$, $c = 1000$), showing the extreme cases of $k \times a \rightarrow 0$ (left) and $k \times a \rightarrow \pi$ (right).}
	\label{fig18a}
\end{figure}

Reflection and transmission coefficients shown in figure \ref{fig18} display very narrow-band one-way reflection properties. High values of $|R_{-}|$ are observed for $k \times a \rightarrow 0$ and $k \times a \rightarrow \pi$, and are not sensitive to the selection of $c$. Close-up views for $k \times a \rightarrow 0$ and $k \times a \rightarrow \pi$ are shown in figure \ref{fig18a}. Note that for $k \times a \rightarrow \pi$ the value of $|R_{-}|$ is close to one, the perfect reflection, while for $k \times a \rightarrow 0$, $|R_{-}|$ converges to a lower value. These surprising results  will be analyzed in detail in  Section \ref{sec5}.

\subsection{Impedances with different passive damping properties.}\label{sec03}

Selecting an arbitrary, but different from considered so far, configuration of passive damping properties of scatterers we assume - analogously to section \ref{pert} - that $m_-''$ and $m_+''$ are related by $m_+'' = c \ m_-''$ and $m_+'' < 0$. The constant must satisfy $c \in (-\infty,-1) \cup (+1,+\infty)$, where values $c = \{-1; +1\}$ recover results obtained in sections \ref{112=} and \ref{113=}, and $c = \infty$ is equivalent to result presented in section \ref{sec02}. The relation between $m_+'$ and $m_-'$ yields
\beq{1008}
\big( K - m_{+}' \cot ka \big) \ c \ \cot ka - m_{-}' = 0, \ \ \ \ \ m_+'' < 0,
\eeq

\begin{figure}[h!]
	\centering
	\begin{subfigure}{}
		\includegraphics[width=0.48\linewidth]{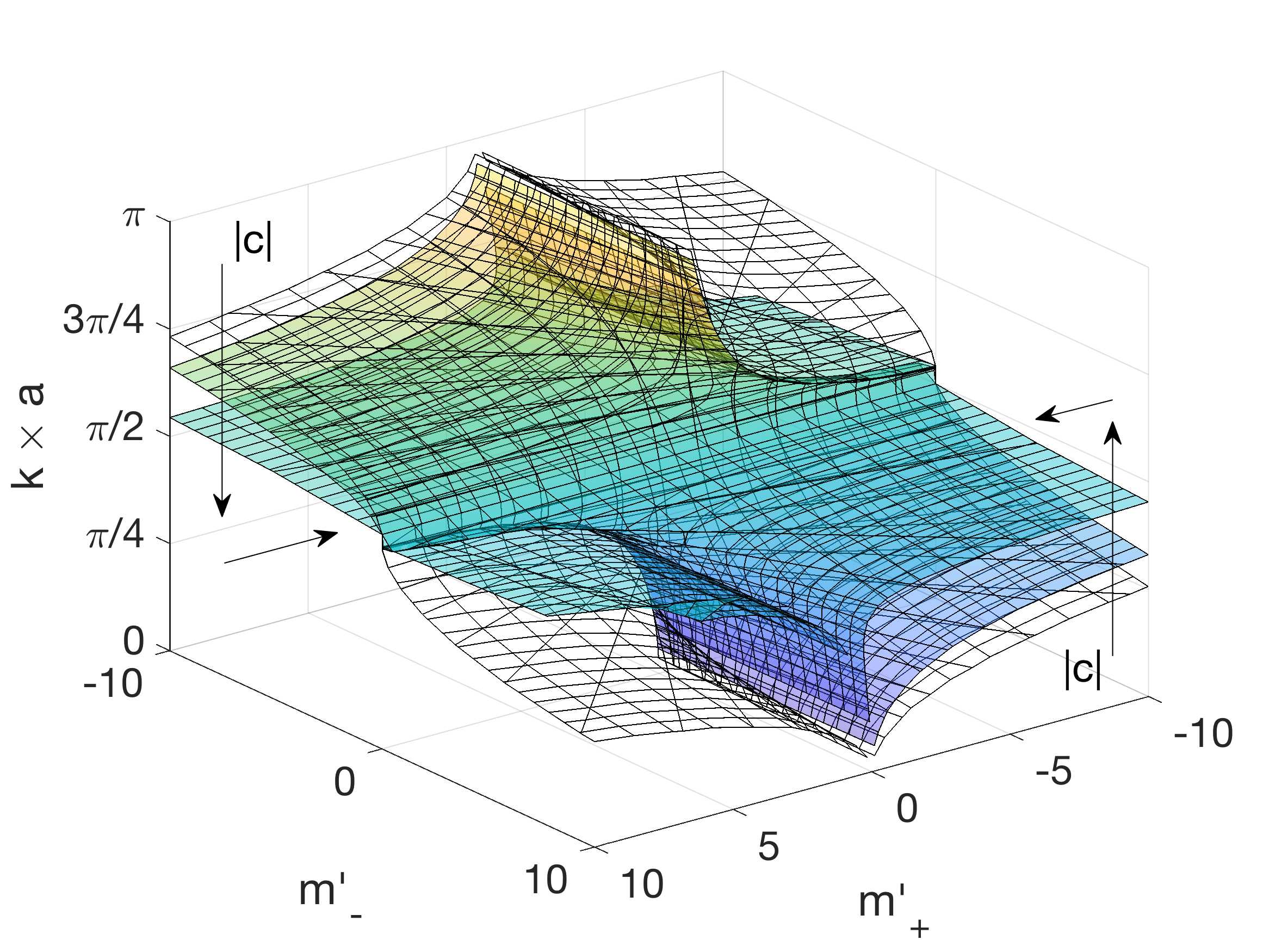}
	\end{subfigure}
	\begin{subfigure}{}
		\includegraphics[width=0.48\linewidth]{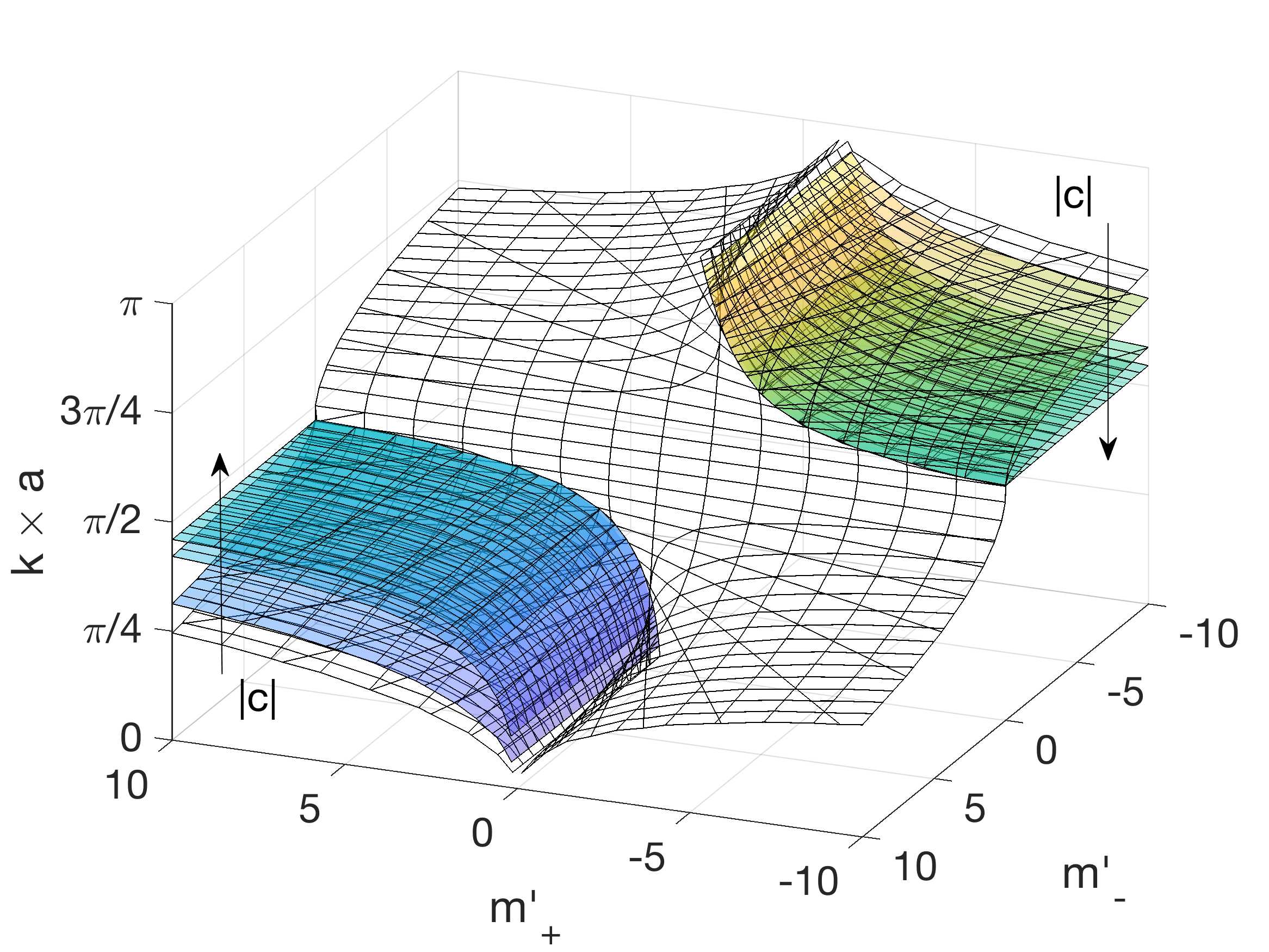}
	\end{subfigure}
	\caption{The design space analogous to that from figure \ref{fig3}, defined by equation \eqref{1008} for $c = \{ \hat{c}; 10 \ \hat{c}; 100 \ \hat{c} \}$, (a), and for $c = \{ -\hat{c}; -10 \ \hat{c}; -100 \ \hat{c} \}$, (b).}
	\label{fig8}
\end{figure}

Selection of $m_+'' < 0$ and $c$ uniquely defines passive damping properties of the two scatterers. Figure \ref{fig8} presents fragments of the design space for $c < -1$ and $c > 1$ for three arbitrarily selected values of $c: |c| = \{\hat{c}; 10 \ \hat{c}; 100 \ \hat{c}\}$, $\hat{c} \geq 1$. It can be seen that for increasingly large values of $|c|$ the shape of the design space converges to that of figure \ref{fig6}, while for small values of $|c|$ to that of figure \ref{fig3}.
Specifically, when $|c|$ is large $m_{-}'' = m_{+}'' \ / \ |c| \rightarrow 0$, resulting in the same damping properties of the scatterers, and $m_{-}' \ / \ |c| \rightarrow 0$ in \eqref{1008}. The latter observations are consistent with results of section \ref{sec02} (see figure \ref{fig6}),  and discussion in \ref{m1im2izero}.

Note that $|m_{-}''| < - m_{+}''$ with $m_{+}''$ and \eqref{-207b} restrict the selection of $m_{-}' > 0$ for $k \times a \in (0, \pi / 2)$ and $m_{-}' < 0$ for $k \times a \in (\pi / 2, \pi)$. The value of $m_{+}'$ is then given by
\beq{1338}
m_{+}' = \big( K - \frac{m_{-}'}{c} \tan ka \big) \tan ka.
\eeq

For negative $|c|$ (figure \ref{fig8}b) $m_{+}'$ behaves as $\sim \tan^2 ka$ for small $|c|$ and as $\sim \tan ka$ for large $|c|$. For small positive $|c|$ (figure \ref{fig8}a) $m_{+}'$ takes values $\sim -\tan^2 ka$ and for large positive $|c|$, $m_{+} \sim \tan ka$. It should be noted that as $m_{+}'$ can change its sign, there exists $m_{+}' = 0$ choice (i.e. equal but opposite real parts of the normalized impedances) and can be achieved by a configuration satisfying
\beq{1339}
m_{-}' = c K \cot ka.
\eeq

It follows from equation \eqref{-207b}  that \eqref{1339} can be satisfied only by $c > 0$ for the system to be passive.

\section{Maximum reflection for $ka\ll 1$ and  $ka\approx \pi$} \label{sec5}

We return  to the surprising results indicating that significant one way reflection  $(|R_-|>0.8)$ is possible for $ka \ll 1$, and that almost unitary one way reflection  $(|R_-|\approx 1)$  can be achieved for $ka \approx \pi$, as illustrated in figures \ref{rtcase3} and 
\ref{fig18a}.    Here we derive analytical expressions that help explain the extreme values of $|R_-|$. These effects are associated with small imaginary parts for the impedances $-m_1'', -m_2'' \ll 1$; we therefore concentrate on the case considered in Section \ref{sec02}
for $m_1'' = m_2'' \equiv m''$,  corresponding to the impedances defined by Eqs.\  \eqref{1900}.
The single parameter $m''$  yields, from  \eqref{88} and \eqref{3-57} a reflection coefficient  
\beq{043}
R_-= \frac{2m'' \cos ka}
	{ \big(	\frac{m''}{\sin ka} 	+\ii g(a) \tan ka \big)^2 	}.
	\eeq
Equation \eqref{043} may be written as 
\beq{045}
R_-= \frac{-\cos ka}
	{ \big(	\frac{1}{2y} 	-\frac{\ii g(a) y}{\cos ka} \big)^2	}
	\ \ \text{where } \ y = \frac{\sin ka}{\sqrt{-2 m''}}.
	\eeq
This form allows us to easily find the asymptotic limits appropriate to the two cases of interest, which are considered next. 
	
\subsection{Low frequency maximal absorption} \label{sec51}

  
Expanding the expression  \eqref{045}   for 
$ka \ll 1$, $-m'' \ll 1$ indicates the preferred scaling $m'' = $O$(ka)^2$.  Thus, 
\beq{-47}
R_- \approx \frac{-4}
{\big(\frac 1{\gamma} + (1+\ii )\gamma  \big)^2}
\ \  \text{with} \ \gamma = \frac{ka}{\sqrt{-2 m''}} .
\eeq
Hence, $|R_-|  = 4 /\big( \frac 1{\gamma^2} + 2\gamma^2 +2\big)$, implying that the  maximum reflection    is $R_-   = 2 (\sqrt{2} -1) e^{\ii 3 \pi/4}$ at  $\gamma =2^{-1/4}$, corresponding to 
\beq{-48}
R_- =  -0.5858 + 0.5858\, \ii  = 0.8284 \, e^{\ii 3 \pi/4} .
\eeq
In summary, for a given value of  $-m'' \ll 1$ the reflection for small $ka$ is given approximately 
by  \eqref{-47}, with maximum amplitude $|R_-|  =0.8248$   at $ka = 1.1892\sqrt{-m''}$.  An example is shown in Figure \ref{asymptotics}. 

 \subsection{Maximal absorption for $ka \approx \pi$} \label{sec52}

\begin{figure}[h!]
	\centering
	\begin{subfigure}{}
		\includegraphics[width=0.48\linewidth]{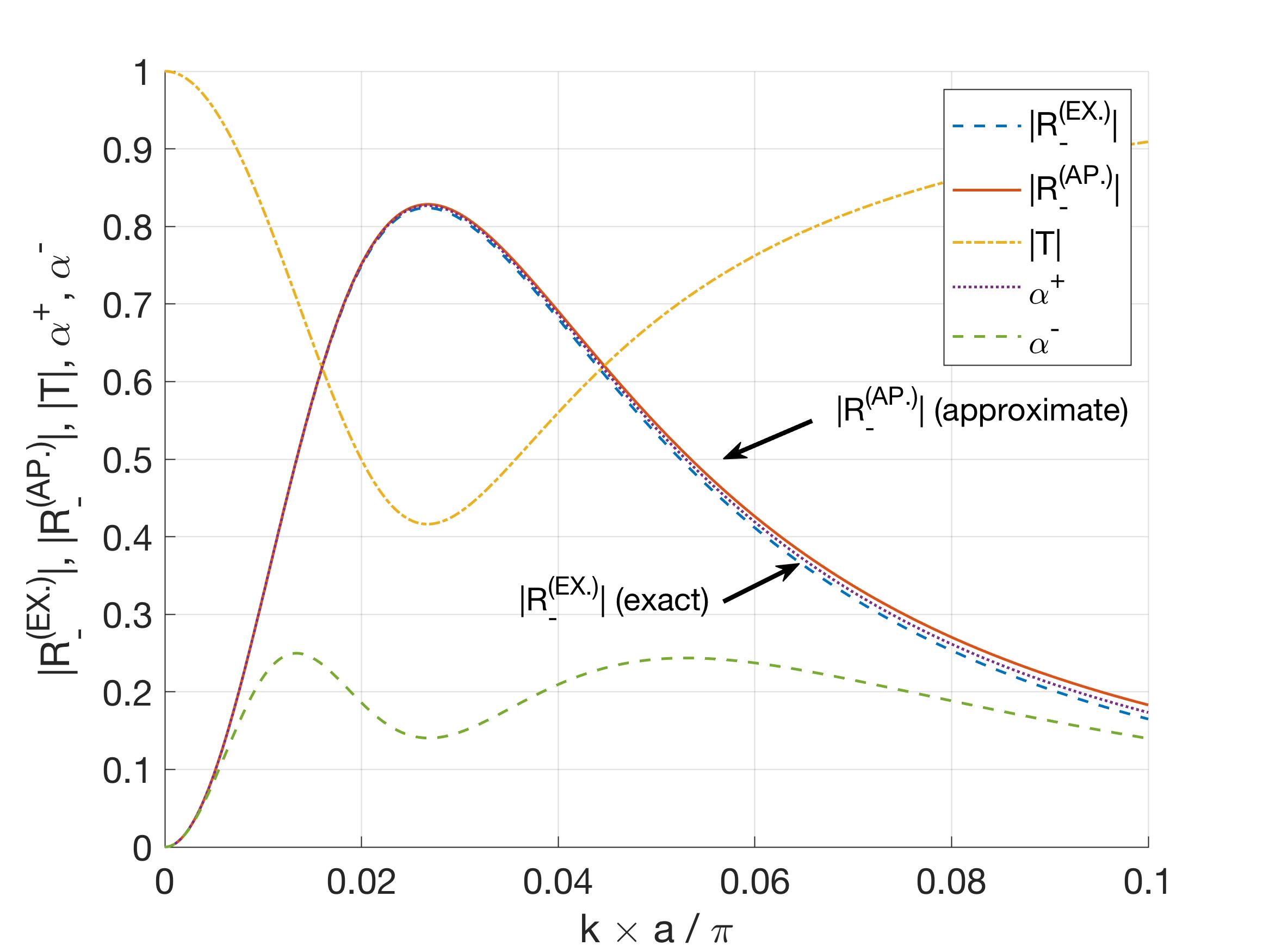}
	\end{subfigure}
	\begin{subfigure}{}
		\includegraphics[width=0.48\linewidth]{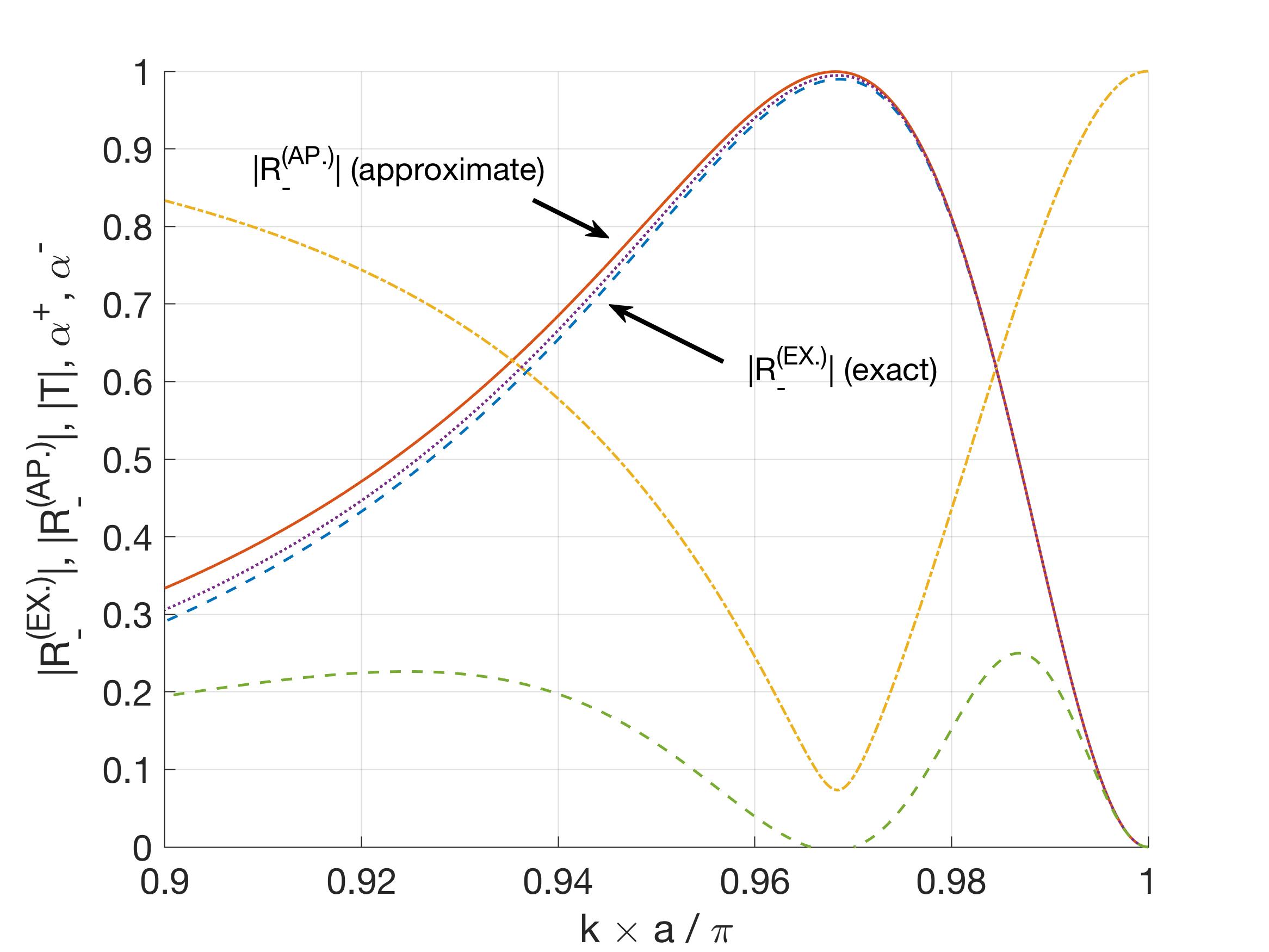}
	\end{subfigure}
	\caption{Reflection coefficient $|R_{-}|$ near   $ka = 0$ (left) and   $ka=\pi$  (right) for  $m_1'' = m_2'' = -0.005$; $|R_{+}| = 0$.  In both cases the exact coefficient is from 
	\eqref{043}.  The approximation for small $ka$ comes from   \eqref{-47}, and  for $ka \approx \pi$  from \eqref{244}. {Absorption coefficients $\alpha^\pm$ computed for approximate values of $R_-$.}}
	\label{asymptotics}
\end{figure}

In this case we assume $\pi - ka \ll 1$, $-m'' \ll 1$.  Expanding the expression  \eqref{045} we find a similar 
 preferred scaling as before, with now $m'' = $O$\big( (\pi - ka)^2 \big)$.  Thus, 
\beq{244}
R_- \approx \frac{4}
{\big(\frac 1{\gamma} + z\gamma  \big)^2} 
\eeq
where  
\beq{9=36}
\gamma = \frac{\pi - ka}{\sqrt{-2 m''}} , \ \  z = 1 - \ii e^{-\pi} = |z| e^{\ii \phi}, 
\eeq
i.e.\  $|z|=1.0009$, $\phi = -0.0432$.  
In this case 
$|R_-|  = 4 /\big( \frac 1{\gamma^2} + |z|^2\gamma^2 +2\big)$, implying that the  maximum reflection  coefficient  is $R_-   = 1/( z \cos^2 \frac{\phi}2 )$ at  $\gamma =|z|^{-1/2}
= 0.9995 $, corresponding to 
\beq{-418}
R_- =  0.9986 + 0.0432 \,\ii  = 0.9995 \, e^{-\ii \phi }.
\eeq
The maximum $|R_-|$ is very close to, but not equal to unity, as illustrated in the example of Figure \ref{asymptotics}. 
In summary, for a given value of  $-m'' \ll 1$ the reflection for $ka \approx \pi$ is given approximately 
by \eqref{244} and \eqref{9=36}, with almost unit maximum amplitude at $ka = \pi - 1.4136 \sqrt{-m''}$

Finally, we note from  Eqs.\  \eqref{1900} that the  impedances for both cases, $ka $ near zero and $\pi$, are 
\beq{1980}
m_{1 \over 2} \approx \pm \frac{\sqrt{-  m''/2}}{\gamma}   +\ii m'' .
\eeq
These correspond to lightly damped oscillators, one being an effective mass, the other a stiffness.  

{
\section{Maximum reflection for $ka>\pi$} \label{sec6}
Finally, we consider the unique value of $m_+ = m_1+m_2$ for which the transmission is identically zero, $T=0$, in addition to $R_+ = 0$.  It follows from 
eq.\ \eqref{8-8} that both $R_-$ and $T$ vanish if 
\beq{711}
m_+ = 2g(0) - 2g(a)\cos ka - \ii 2 \sin^2 ka, 
\eeq
implying, using eq.\ \eqref{92}, that 
\beq{714}
\begin{aligned}
m_1 &= -\frac 12  + \big(\sin ka + \frac 12 e^{-ka}\big) e^{-\ii ka}, 
\\
m_2 &= -\frac 12  +   \frac 12 e^{-ka}  e^{\ii ka} ,
\\
R_- &=  -e^{-\ii ka} + \ii e^{-ka}  .
\end{aligned}
\eeq
The lowest value of $ka$ for which both impedances are passive $m_1'' , m_2'' \le 0$ is $ka = 1.0067 \, \pi$, and they remain passive until $ka = 2\pi$.  Therefore, for almost all of the range $ka = (\pi,2\pi)$ these impedances yield $R_+=T=0$ with 
almost unit amplitude one way  reflection $|R_-|>0.986$.   The damping of attachment 2 is very small, $m_2'' > -0.007$, with almost all the damping in resonator 1. 
}

\section{Summary and conclusions}  \label{sec7}

We have presented for the first time a flexural wave analogue of the one-way absorption effect  in acoustics.  Similar to the acoustic setup, we consider a pair of adjacent lumped elements which breaks the symmetry of the scattering, in this case  oscillators
attached pointwise on a beam. It is found that at least one of the oscillators must be damped in order to have zero reflection from one direction with significant reflection from the opposite direction.  The method of analysis developed in Section \ref{sec2} can be easily generalized to handle larger clusters of oscillators; however,  we have shown that two are sufficient to achieve one-way absorption. 

The starting point for finding possible combinations of oscillator pairs is Eq.\ \eqref{4-3} which guarantees one reflection coefficient vanishes (in this case $R_+=0$).  This condition provides a single relation between  the normalized impedances $m_1 = m_1' + \ii m_1''$ and $m_2 = m_2' + \ii m_2''$, which are otherwise unconstrained except that they must correspond to passive oscillators $(m_1'' \le 0,m_2'' \le 0)$. This leaves a large set of possible configurations that may be considered.  The bulk of the paper, Sections \ref{sec4} and \ref{sec5}, is devoted to 
 investigating this  large space of  design parameters. 
The numerous examples demonstrate  that  one way  absorption can be realized by various configurations of the scatterers and their mechanical properties, e.g. a combination of a single damper with two mass-spring elements, a damper with a single mass-spring oscillator, a pinned point with a damper or combinations of two dampers with oscillators. 

The examples discussed in Section \ref{sec4} and summarized in Table \ref{table1} indicate that significant one-way reflection can be obtained for attachments spaced less than $\lambda / 2$ apart. For instance, if one of the impedances is real, corresponding to a mass or a stiffness,  and the other attachment is a pure damper, 
then almost unit reflection can be achieved for an  approximate spacing of  $\lambda / 2$ (see Figure \ref{fig4}). Alternatively, if one of the points is pinned and the other attachment is a damped oscillator then relatively broadband and significant one-way reflection is possible, as shown in Figure \ref{fig5}. {This finding for flexural waves differs substantially from the acoustic case.  Here perfect one-way absorption can be obtained for a combination of a damped oscillator with a pinned point - and is attributed to the partial transfer of elastic waves through the pinned (i.e.\ $w = 0$) point due to rotations of the beam cross sections.} We find that virtually perfect one-way absorption is possible with two attachments with the same but small damping, if the spacing between them is slightly less than $\lambda / 2$. {Such a configuration requires  real parts of the scatterers' impedances of opposite signs (see \eqref{1980}) and could be achieved by properly selected high values of mass (positive sign) and stiffness (negative sign) parameters for each of them (see \eqref{-1}).}  This effect, shown in Figure \ref{asymptotics} has also been verified using asymptotics based on the small parameter 
$m_1''=m_2''$.   Surprisingly,  the same setup of  two attachments with equal and  small damping yields a reflectivity  of magnitude $0.82$ for very small spacing $a \ll \lambda$, also shown in  Figure \ref{asymptotics}. The strictly sub-wavelength nature of this  effect means that the pair of damped oscillators may be viewed as a single attachment, i.e.\  a flexural wave  Willis element \cite{Muhlestein2017a}.

The results in this paper open up  new possibilities in structural wave dynamics.  For instance, one could in principle design vibration absorbers that are not only frequency selective, but also depend on where the noise is incident from.  In this paper we have shown that a large design space exists; however, there is more work to be done in interpreting this type of phenomenon in terms of realistic adaptive oscillators.  This requires mapping the non-dimensional impedances found back to realistic oscillator dynamics, as in the spring-mass-damper models of Eq.\ \eqref{-1}.  The present results use only translational impedances (point forces) in the context of the classical beam theory, but could be extended to include concentrated moments and more refined engineering theories.  
For instance,  further analysis of the results in Section \ref{sec51} in which the point attachments are very close together  would benefit from a more precise theory, such as  Timoshenko's, that better models the near-field of concentrated forces.    



\section*{Acknowledgments}
The work of ANN was  supported by the National Science
Foundation under Award No. EFRI 1641078 and  the Office of
Naval Research under MURI Grant No. N00014-13-1-0631.

PP acknowledges support from the National Centre for Research and Development under the research programme LIDER (Project No. LIDER/317/L-6/ 14/NCBR/2015).


\begin{thebibliography}{10}

\bibitem{Muhlestein2017a}
Michael~B. Muhlestein, Caleb~F. Sieck, Preston~S. Wilson, and Michael~R.
  Haberman.
\newblock Experimental evidence of {W}illis coupling in a one-dimensional
  effective material element.
\newblock {\em Nature Comm.}, 8:15625, Jun 2017.

\bibitem{Su2018a}
Xiaoshi Su and Andrew~N. Norris.
\newblock Retrieval method for the bianisotropic polarizability tensor of
  {W}illis acoustic scatterers.
\newblock {\em Physical Review B}, 98(17), Nov 2018.

\bibitem{Merkel2018}
Aur{\'{e}}lien Merkel, Vicent Romero-Garci\'a, Jean-Philippe Groby, Jensen Li,
  and Johan Christensen.
\newblock Unidirectional zero sonic reflection in passive {PT} -symmetric
  {W}illis media.
\newblock {\em Physical Review B}, 98(20), nov 2018.

\bibitem{Mead82}
D.~J. Mead.
\newblock {\em Structural wave motion}, chapter 9 in Noise and Vibration, R. G.
  White and J. G. Walker (editors), pages 207--226.
\newblock Ellis Horwood Publishers, Chichester, 1982.

\bibitem{Brennan1997}
M~J Brennan.
\newblock Vibration control using a tunable vibration neutralizer.
\newblock {\em Proceedings of the Institution of Mechanical Engineers, Part C:
  Journal of Mechanical Engineering Science}, 211(2):91--108, feb 1997.

\bibitem{Brennan1999}
M.J. Brennan.
\newblock Control of flexural waves on a beam using a tunable vibration
  neutraliser.
\newblock {\em Journal of Sound and Vibration}, 222(3):389--407, may 1999.

\bibitem{El-Khatib2005}
H.M. El-Khatib, B.R. Mace, and M.J. Brennan.
\newblock Suppression of bending waves in a beam using a tuned vibration
  absorber.
\newblock {\em Journal of Sound and Vibration}, 288(4-5):1157--1175, dec 2005.

\bibitem{Yang2016}
Cheng Yang and Li~Cheng.
\newblock Suppression of bending waves in a beam using resonators with
  different separation lengths.
\newblock {\em Journal of the Acoustical Society of America},
  139(5):2361--2371, may 2016.

\bibitem{Torrent2013}
Daniel Torrent, Didier Mayou, and Jos{\'{e}} S{\'{a}}nchez-Dehesa.
\newblock Elastic analog of graphene: Dirac cones and edge states for flexural
  waves in thin plates.
\newblock {\em Physical Review B}, 87(11), mar 2013.

\bibitem{Evans2007}
D.~V. Evans and R.~Porter.
\newblock Penetration of flexural waves through a periodically constrained thin
  elastic plate in vacuo and floating on water.
\newblock {\em Journal of Engineering Mathematics}, 58(1):317--337, Aug 2007.

\end{thebibliography}
\end{document}